\documentclass[11pt]{article}
\usepackage[dvips]{graphicx}
\usepackage{float}
\usepackage{epsfig}
\usepackage{latexsym,amsmath,amsfonts,amssymb}
\usepackage{rotating}
\usepackage[american]{babel}
\usepackage[dvips]{graphicx}
\usepackage[section]{placeins}
\usepackage{bbm}
\usepackage{color}
\usepackage[frozencache,cachedir=minted-cache]{minted}
\usepackage{slashed}
\usepackage[unicode]{hyperref}
\usepackage{lscape}
\usepackage{bigints}
\usepackage{enumerate}
\usepackage[shortlabels]{enumitem}
\usepackage{subcaption}
\usepackage{tikz}
\usetikzlibrary{decorations.pathreplacing}
\usetikzlibrary{shapes}
\usepackage{soul}
\usepackage{graphicx}
\usepackage{epsfig}
\usepackage{rotating}
\usepackage{amssymb}
\usepackage{dsfont}
\usepackage{psfrag}
\usepackage{amsmath,euscript,array,mathrsfs,amsfonts}
\usepackage{slashed}
\usepackage{array}
\usepackage{youngtab}
\usepackage{color}
\usepackage{bbold}
\usepackage{hyperref}
\hypersetup{
colorlinks = true,
linkcolor = red,
linktocpage = true,
citecolor = blue
}

\usepackage{tikz}
\usetikzlibrary{calc}
 \usetikzlibrary{decorations.text}
 \usetikzlibrary{shapes}

 \usetikzlibrary{decorations.pathmorphing}
\usetikzlibrary{decorations.pathreplacing}
\usetikzlibrary{arrows.meta}
\tikzset{
 >={To[length=5pt]}
 }
\usetikzlibrary{shapes, shapes.geometric, shapes.symbols, shapes.arrows, shapes.multipart, shapes.callouts, shapes.misc}
\tikzset{snake it/.style={decorate, decoration=snake}}
\tikzset{7brane/.style={circle, draw=black, fill=black,ultra thick,inner sep=1.5 pt, minimum size=1 pt,}, c/.default={4pt}}
\tikzset{cross/.style={cross out, draw=black,thick, minimum size=2*(#1-\pgflinewidth), inner sep=0pt, outer sep=0pt}, cross/.default={5pt}}
\tikzset{big7brane/.style={circle, draw=black, fill=black,ultra thick,inner sep=2.5 pt, minimum size=1 pt,}, c/.default={4pt}}
\tikzset{u/.style={circle, draw=black, fill=white,inner sep=2 pt, minimum size=2 pt,},f/.style={square, draw=black, fill=white,ultra thick,inner sep=4 pt, minimum size=2 pt,}}
\tikzset{so/.style={circle, draw=black, fill=red,inner sep=2 pt, minimum size=2 pt,},f/.style={square, draw=black, fill=white,ultra thick,inner sep=4 pt, minimum size=2 pt,}}
\tikzset{sp/.style={circle, draw=black, fill=blue,inner sep=2 pt, minimum size=2 pt,},f/.style={square, draw=black, fill=white,ultra thick,inner sep=4 pt, minimum size=2 pt,}}
\tikzset{uf/.style={rectangle, draw=black, fill=white,inner sep=3 pt, minimum size=4 pt,}}
\tikzset{spf/.style={rectangle, draw=black, fill=blue, thick,inner sep=3 pt, minimum size=4 pt, circle, draw=black, fill=blue,thick,inner sep=2 pt, minimum size=2 pt,},f/.style={square, draw=black, fill=white,ultra thick,inner sep=4 pt, minimum size=2 pt,}}
\tikzset{sof/.style={rectangle, draw=black, fill=red, thick,inner sep=3 pt, minimum size=4 pt,}}
\usetikzlibrary{positioning}
\usetikzlibrary{arrows}
\usetikzlibrary{decorations.pathreplacing}
\usetikzlibrary{shapes}

\makeatletter\def\l@subsubsection#1#2{}%
\makeatother

\renewcommand\theequation{\arabic{section}.\arabic{equation}} 

\def\cA{{\cal A}}

\def\cF{{\cal F}}

\def\CC{\ensuremath{\mathds C}}
\def\RR{\ensuremath{\mathds R}}
\def\ZZ{\ensuremath{\mathds Z}}

\DeclareMathOperator{\vol}{vol}
\DeclareMathOperator{\Vol}{Vol}
\DeclareMathOperator{\sech}{sech}
\DeclareMathOperator{\tr}{tr}

\DeclareMathOperator{\csch}{csch}

\newcommand{\be}{\begin{equation}}
\newcommand{\ee}{\end{equation}}
\newcommand{\ba}{\begin{array}}
\newcommand{\ea}{\end{array}}

\newcommand{\textitle}[1]{\texorpdfstring{#1}{}}

\def\im{Invent. Math.}

\def\hat{\widehat}
\def\a{\alpha}
\def\b{\beta}
\def\c{\gamma}
\def\d{\delta}
\def\f{\phi} 
\def\vf{\varphi} 
\def\tvf{\tilde{\varphi}}
\def\vp{\varphi}
\def\g{\gamma}
\def\h{\eta}
\def\j{\psi}
\def\k{\kappa} 
\def\l{\lambda}
\def\m{\mu}
\def\n{\nu}
\def\o{\omega} \def\w{\omega}
\def\p{\pi}
\def\q{\theta} \def\th{\theta} 
\def\r{\rho} 
\def\s{\sigma} 
\def\t{\tau}
\def\u{\upsilon}
\def\x{\xi}
\def\z{\zeta}
\def\pt{\tilde{\varphi}}
\def\tt{\tilde{\theta}}
\def\lab{\label}
\def\6{\partial}
\def\wg{\wedge}
\def\bpsi{\bar{\psi}}
\def\bt{\bar{\theta}}
\def\bvf{\bar{\varphi}}
\def\W{\Omega}

\newcommand{\td}{\mathrm{d}}

\DeclareMathOperator{\str}{str}

\newcommand{\beq}{\begin{equation}}
\newcommand{\eeq}{\end{equation}}
\newcommand{\bea}{\begin{eqnarray}}
\newcommand{\eea}{\end{eqnarray}}

\newcommand{\beqs}{\begin{eqnarray}}
\newcommand{\eeqs}{\end{eqnarray}}
\newcommand{\bal}{\begin{aligned}}
\newcommand{\eal}{\end{aligned}}
\makeatletter
\newcommand\setItemnumber[1]{\setcounter{enum\romannumeral\@enumdepth}{\numexpr#1-1\relax}}
\makeatother
\begin{document}
\baselineskip=15.5pt
\pagestyle{plain}
\setcounter{page}{1}

\def\del{{\partial}}
\def\vev#1{\left\langle #1 \right\rangle}
\def\cn{{\cal N}}
\def\co{{\cal O}}


\def\IC{{\mathbb C}}
\def\IR{{\mathbb R}}
\def\IZ{{\mathbb Z}}
\def\RP{{\bf RP}}
\def\CP{{\bf CP}}
\def\Poincaré{{Poincar\'e }}
\def\tr{{\rm tr}}
\def\tp{{\tilde \Phi}}

\def\TL{\hfil$\displaystyle{##}$}
\def\TR{$\displaystyle{{}##}$\hfil}
\def\TC{\hfil$\displaystyle{##}$\hfil}
\def\TT{\hbox{##}}
\def\HLINE{\noalign{\vskip1\jot}\hline\noalign{\vskip1\jot}}
\def\seqalign#1#2{\vcenter{\openup1\jot
 \halign{\strut #1\cr #2 \cr}}}
\def\lbldef#1#2{\expandafter\gdef\csname #1\endcsname {#2}}
\def\eqn#1#2{\lbldef{#1}{(\ref{#1})}%
\begin{equation} #2 \label{#1} \end{equation}}
\def\eqalign#1{\vcenter{\openup1\jot
 \halign{\strut\span\TL & \span\TR\cr #1 \cr
 }}}

\def\eno#1{(\ref{#1})}
\def\href#1#2{#2}
\def\half{\frac{1}{2}}



\def\ads{{\it AdS}}
\def\adsp{{\it AdS}$_{p+2}$}
\def\cft{{\it CFT}}

\newcommand{\ber}{\begin{eqnarray}}
\newcommand{\eer}{\end{eqnarray}}

\newcommand{\beqar}{\begin{eqnarray}}
\newcommand{\cO}{{\cal O}}
\newcommand{\cT}{{\cal T}}
\newcommand{\cR}{{\cal R}}
\newcommand{\eeqar}{\end{eqnarray}}
\newcommand{\tht}{\thteta}
\newcommand{\lm}{\lambda}\newcommand{\Lm}{\Lambda}


\newcommand{\nonu}{\nonumber}
\newcommand{\oh}{\displaystyle{\frac{1}{2}}}
\newcommand{\dsl}
 {\kern.06em\hbox{\raise.15ex\hbox{$/$}\kern-.56em\hbox{$\partial$}}}
\newcommand{\as}{\not\!\! A}
\newcommand{\ps}{\not\! p}
\newcommand{\ks}{\not\! k}
\newcommand{\D}{{\cal{D}}}
\newcommand{\dv}{d^2x}
\newcommand{\Z}{{\cal Z}}
\newcommand{\N}{{\cal N}}
\newcommand{\Dsl}{\not\!\! D}
\newcommand{\Bsl}{\not\!\! B}
\newcommand{\Psl}{\not\!\! P}

\newcommand{\eeqarr}{\end{eqnarray}}


\def\del{{\delta^{\hbox{\sevenrm B}}}} \def\ex{{\hbox{\rm e}}}
\def\azb{A_{\bar z}} \def\az{A_z} \def\bzb{B_{\bar z}} \def\bz{B_z}
\def\czb{C_{\bar z}} \def\cz{C_z} \def\dzb{D_{\bar z}} \def\dz{D_z}
\def\im{{\hbox{\rm Im}}} \def\mod{{\hbox{\rm mod}}} \def\tr{{\hbox{\rm Tr}}}
\def\ch{{\hbox{\rm ch}}} \def\imp{{\hbox{\sevenrm Im}}}
\def\trp{{\hbox{\sevenrm Tr}}} \def\vol{{\hbox{\rm Vol}}}
\def\rl{\Lambda_{\hbox{\sevenrm R}}} \def\wl{\Lambda_{\hbox{\sevenrm W}}}
\def\fc{{\cal F}_{k+\cox}} \def\vev{vacuum expectation value}
\def\nodiv{\mid{\hbox{\hskip-7.8pt/}}}
\def\ie{{\em i.e.}}
\def\ie{\hbox{\it i.e.}}

\def\CC{{\mathchoice
{\rm C\mkern-8mu\vrule height1.45ex depth-.05ex
width.05em\mkern9mu\kern-.05em}
{\rm C\mkern-8mu\vrule height1.45ex depth-.05ex
width.05em\mkern9mu\kern-.05em}
{\rm C\mkern-8mu\vrule height1ex depth-.07ex
width.035em\mkern9mu\kern-.035em}
{\rm C\mkern-8mu\vrule height.65ex depth-.1ex
width.025em\mkern8mu\kern-.025em}}}

\def\RR{{\rm I\kern-1.6pt {\rm R}}}
\def\NN{{\rm I\!N}}
\def\ZZ{{\rm Z}\kern-3.8pt {\rm Z} \kern2pt}
\def\IB{\relax{\rm I\kern-.18em B}}
\def\ID{\relax{\rm I\kern-.18em D}}
\def\II{\relax{\rm I\kern-.18em I}}
\def\IP{\relax{\rm I\kern-.18em P}}
\newcommand{\CS}{{\scriptstyle {\rm CS}}}
\newcommand{\CSs}{{\scriptscriptstyle {\rm CS}}}
\newcommand{\rc}{\nonumber\\}
\newcommand{\bear}{\begin{eqnarray}}
\newcommand{\eear}{\end{eqnarray}}

\newcommand{\LL}{{\cal L}}

\def\mani{{\cal M}}
\def\calo{{\cal O}}
\def\calb{{\cal B}}
\def\calw{{\cal W}}
\def\calz{{\cal Z}}
\def\cald{{\cal D}}
\def\calc{{\cal C}}
\newcommand{\gt}{\tilde{g}}

\def\to{\rightarrow}
\def\ele{{\hbox{\sevenrm L}}}
\def\ere{{\hbox{\sevenrm R}}}
\def\zb{{\bar z}}
\def\wb{{\bar w}}
\def\nodiv{\mid{\hbox{\hskip-7.8pt/}}}
\def\menos{\hbox{\hskip-2.9pt}}
\def\dr{\dot R_}
\def\drr{\dot r_}
\def\ds{\dot s_}
\def\da{\dot A_}
\def\dga{\dot \gamma_}
\def\ga{\gamma_}
\def\dal{\dot\alpha_}
\def\al{\alpha_}
\def\cl{{closed}}
\def\cls{{closing}}
\def\vev{vacuum expectation value}
\def\tr{{\rm Tr}}
\def\to{\rightarrow}
\def\too{\longrightarrow}

\newcommand{\dd}{\mathrm{d}}

\def\a{\alpha}
\def\b{\beta}
\def\c{\gamma}
\def\d{\delta}
\def\e{\epsilon} 
\def\F{\Phi}
\def\f{\phi} 
\def\vf{\varphi} \def\tvf{\tilde{\varphi}}
\def\vp{\varphi}
\def\g{\gamma}
\def\h{\eta}
\def\j{\psi}
\def\k{\kappa} 
\def\l{\lambda}
\def\m{\mu}
\def\n{\nu}
\def\o{\omega} \def\w{\omega}
\def\q{\theta} \def\th{\theta} 
\def\r{\rho} 
\def\s{\sigma} 
\def\t{\tau}
\def\u{\upsilon}
\def\x{\xi}
\def\X{\Xi}
\def\z{\zeta}
\def\pt{\tilde{\varphi}}
\def\tt{\tilde{\theta}}
\def\lab{\label}
\def\6{\partial}
\def\wg{\wedge}
\def\atanh{{\rm arctanh}}
\def\bpsi{\bar{\psi}}
\def\bt{\bar{\theta}}
\def\bvf{\bar{\varphi}}

\def\ft#1#2{{\textstyle{{\scriptstyle #1}\over {\scriptstyle #2}}}}
\def\fft#1#2{{#1 \over #2}}
\def\del{\partial}
\def\sst#1{{\scriptscriptstyle #1}}

\def\dalemb#1#2{{\vbox{\hrule height .#2pt
 \hbox{\vrule width.#2pt height#1pt \kern#1pt
 \vrule width.#2pt}
 \hrule height.#2pt}}}
\def\square{\mathord{\dalemb{6.8}{7}\hbox{\hskip1pt}}}
\def\hF{\hat F}
\def\tA{\widetilde A}
\def\tcA{{\widetilde{\cal A}}}
\def\tcF{{\widetilde{\cal F}}}
\def\hA{\hat{\cal A}}
\def\cF{{\cal F}}
\def\cA{{\cal A}}
\def\wdg{{\sst \wedge}}

\def\0{{\sst{(0)}}}
\def\1{{\sst{(1)}}}
\def\2{{\sst{(2)}}}
\def\3{{\sst{(3)}}}
\def\4{{\sst{(4)}}}
\def\5{{\sst{(5)}}}
\def\6{{\sst{(6)}}}
\def\7{{\sst{(7)}}}
\def\8{{\sst{(8)}}}
\def\n{{\sst{(n)}}}
\def\tV{\widetilde V}
\def\tW{\widetilde W}
\def\tH{\widetilde H}
\def\tE{\widetilde E}
\def\tF{\widetilde F}
\def\tA{\widetilde A}
\def\tP{{\widetilde P}}
\def\tD{\widetilde D}
\def\bA{\bar{\cal A}}
\def\bF{\bar{\cal F}}
\def\tG{\widetilde G}
\def\tT{\widetilde T}
\def\Z{\rlap{\sf Z}\mkern3mu{\sf Z}}
\def\R{\rlap{\rm I}\mkern3mu{\rm R}}
\def\G{{\cal G}}
\def\gg{\bf g}
\def\CS{{\cal S}}
\def\S{{\cal S}}
\def\P{{\cal P}}
\def\ep{\epsilon}
\def\td{\tilde}
\def\wtd{\widetilde}
\def\half{{\textstyle{1\over2}}}
\def\Qw{{Q_{\rm wave}}}
\def\Qnut{{Q_{\sst{\rm NUT}}}}
\def\mun{{\mu_{\sst{\rm NUT}}}}
\def\muw{{\mu_{\rm wave}}}
\let\a=\alpha \let\b=\beta \let\g=\gamma \let\d=\delta \let\e=\epsilon
\let\z=\zeta \let\h=\eta \let\q=\theta \let\i=\iota \let\k=\kappa
\let\l=\lambda \let\m=\mu \let\n=\nu \let\x=\xi \let\p=\pi \let\r=\rho
\let\s=\sigma \let\t=\tau \let\u=\upsilon \let\f=\phi \let\c=\chi \let\y=\psi
\let\w=\omega \let\D=\Delta \let\Q=\Theta \let\L=\Lambda
\let\X=\Xi \let\U=\Upsilon \let\F=\Phi \let\Y=\Psi
\let\C=\Chi \let\W=\Omega 
\let\la=\label \let\ci=\cite \let\re=\ref
\let\se=\section \let\sse=\subsection \let\ssse=\subsubsection 
\def\bd{\begin{document}} \def\ed{\end{document}}
\def\ds{\documentstyle} \let\fr=\frac \let\bl=\bigl \let\br=\bigr
\let\Br=\Bigr \let\Bl=\Bigl 
\let\bm=\bibitem
\let\na=\nabla
\let\pa=\partial \let\ov=\overline 
\def\ba{\begin{eqnarray}}
\def\ea{\end{eqnarray}}
\def\ft#1#2{{\textstyle{{\scriptstyle #1}\over {\scriptstyle #2}}}}
\def\fft#1#2{{#1 \over #2}}
\def\del{\partial}
\def\sst#1{{\scriptscriptstyle #1}}
\def\oneone{\rlap 1\mkern4mu{\rm l}}
\def\ie{{\it i.e.\ }}
\def\via{{\it via}}
\def\semi{{\ltimes}}
\def\str{{\rm str}}
\def\jm{{\rm j}}
\def\im{{\rm i}}
\def\mapright#1{\smash{\mathop{-\!\!\!-\!\!\!-\!\!\!-\!\!\!-\!\!\!
 \longrightarrow}\limits^{#1}}}
\def\maprightt#1#2{\smash{\mathop{-\!\!\!-\!\!\!-\!\!\!-\!\!\!-\!\!\!
 \longrightarrow}\limits^{#1}_{#2}}}

\newcommand{\ho}[1]{$\, ^{#1}$}
\newcommand{\hoch}[1]{$\, ^{#1}$}
\newcommand{\ra}{\rightarrow}
\newcommand{\lra}{\longrightarrow}
\newcommand{\Lra}{\Leftrightarrow}
\newcommand{\bp}{\tilde \beta^\prime}
\newcommand{\Tr}{{\rm Tr} } 
\def\rme{{\rm e}}


\newfont{\namefont}{cmr10}
\newfont{\addfont}{cmti7 scaled 1440}
\newfont{\boldmathfont}{cmbx10}
\newfont{\headfontb}{cmbx10 scaled 1728}





\newcommand{\hyph}[1]{$#1$\nobreakdash-\hspace{0pt}}
\providecommand{\abs}[1]{\lvert#1\rvert}
\newcommand{\Nugual}[1]{$\mathcal{N}= #1 $}
\newcommand{\sub}[2]{#1_\text{#2}}
\newcommand{\partfrac}[2]{\frac{\partial #1}{\partial #2}}
\newcommand{\bsp}[1]{\begin{equation} \begin{split} #1 \end{split} \end{equation}}
\newcommand{\calF}{\mathcal{F}}
\newcommand{\calO}{\mathcal{O}}
\newcommand{\calM}{\mathcal{M}}
\newcommand{\calV}{\mathcal{V}}
\newcommand{\bbZ}{\mathbb{Z}}
\newcommand{\bbC}{\mathbb{C}}
\newcommand{\cK}{{\cal K}}

\newcommand{\Thq}{\Theta\left(\r-\r_q\right)}
\newcommand{\Dq}{\d\left(\r-\r_q\right)}
\newcommand{\kten}{\kappa^2_{\left(10\right)}}
\newcommand{\pbi}[1]{\imath^*\left(#1\right)}
\newcommand{\tth}{\tilde{\th}}
\newcommand{\tf}{\tilde{\f}}
\newcommand{\tj}{\tilde{\j}}
\newcommand{\tw}{\tilde{\omega}}
\newcommand{\tz}{\tilde{z}}
\newcommand{\prj}[2]{(\partial_r{#1})(\partial_{\j}{#2})-(\partial_r{#2})(\partial_{\j}{#1})}
\def\atanh{{\rm arctanh}}
\def\sech{{\rm sech}}
\def\csch{{\rm csch}}
\allowdisplaybreaks[1]

\def\red{\textcolor[rgb]{0.98,0.00,0.00}}

\newcommand{\Dan}[1] {{\textcolor{blue}{#1}}}

\numberwithin{equation}{section}



%

%
\setcounter{footnote}{0}
\renewcommand{\theequation}{{\rm\thesection.\arabic{equation}}}

\begin{titlepage}

\begin{center}

\vskip .5in 
\noindent

{\Large \bf{Confinement and screening via holographic Wilson loops} }
\bigskip\medskip

Mauro Giliberti$^\dagger$\footnote{ mauro.giliberti@unifi.it}, Ali Fatemiabhari$^*$\footnote{a.fatemiabhari.2127756@swansea.ac.uk} and Carlos Nunez$^*$\footnote{c.nunez@swansea.ac.uk} \\

\bigskip\medskip
{\small 
$^\dagger$ Dipartimento di Fisica e Astronomia, Università degli Studi di Firenze;\\Via G. Sansone 1; I-50019 Sesto Fiorentino (Firenze), Italy.
\\
$^*$ Department of Physics, Swansea University, ~Swansea SA2 8PP, United Kingdom}

\vskip .5cm 
\vskip .9cm 
 	{\bf Abstract }\vskip .1in
\end{center}

\noindent
We present the holographic dual to a family of ${\cal N}=1$ SCFTs in four dimensions, deformed by a VEV leading to a gapped system. We calculate Wilson loops in this system containing adjoint, bifundamental and fundamental matter. We calculate the quark-antiquark energy $E$ in terms of their separation $L$, finding an approximate analytic expression for $E(L)$. This expression shows the transition between conformal, confining and screened behaviours. Interesting phenomenology is discussed in a variety of examples. The tool used is the minimization of the F1 string action, for which the code used is made publicly available. \noindent
\vskip .5cm
\vskip .5cm
\vfill
\eject

\end{titlepage}

\setcounter{footnote}{0}

\small{
\tableofcontents}

\normalsize

\newpage
\renewcommand{\theequation}{{\rm\thesection.\arabic{equation}}}
%
\section{Introduction}
The Maldacena conjecture and its refinements \cite{Maldacena:1997re,Gubser:1998bc,Witten:1998qj} naturally lead to the application of holography to study non-conformal field theories at strong coupling, see for example the early seminal works \cite{Itzhaki:1998dd}, \cite{Witten:1998zw}, \cite{Boonstra:1998mp}, \cite{Girardello:1999hj}, \cite{Polchinski:2000uf} . 

Naturally, these developments led to the holographic study of confining field theories. Roughly, there are two types of constructions for duals to confining QFTs. The first of these approaches uses wrapped branes -- see for example \cite{Witten:1998zw,Maldacena:2000yy,Atiyah:2000zz,Edelstein:2001pu,Maldacena:2001pb}. The second describes the dynamics of a particular two-node quiver field theory, quasi-marginally deformed. On the string side, this is achieved by studying the dynamics of D3 and D5 branes on the conifold \cite{Klebanov:1998hh,Klebanov:2000nc,Klebanov:2000hb,Gubser:2004qj}. It is possible to connect these two lines of study \cite{Maldacena:2009mw,Gaillard:2010qg, Caceres:2011zn, Elander:2011mh}.


Introducing the dynamics of degrees of freedom transforming in the fundamental representation of the gauge group (quarks), is a technically challenging problem. Important progress was achieved in various works. See \cite{Casero:2006pt, Paredes:2006wb, Burrington:2007qd, Casero:2007jj, Bigazzi:2008gd, Hoyos-Badajoz:2008znk, Bigazzi:2008ie, Bigazzi:2008qq, Bigazzi:2009bk, Nunez:2010sf,Benini:2006hh,Benini:2007gx, Bigazzi:2014qsa, Bigazzi:2011it, Bea:2013jxa} for a sample of representative papers. An important feature of these constructions is that the flavour branes (source branes on the gravity side of the duality) are either extended or smeared across all the internal space. As a consequence, instead of working with sharply defined $SU(N_f)$ flavour groups, the constructions above discuss the QFT for which VEVs have broken $SU(N_f)\to U(1)^{N_f}$. This has effects on the dynamics: among them, the fact that for massless fundamental matter the small-$r$ region of the background (corresponding with the IR of the QFT) is
singular, hence calculations close to this region are not trustable. Another consequence is that the phenomenon of screening is slightly obscured: indeed, the screening (the snapping of one `connected' string probe into two `disconnected' ones) is weighted by factors of $g_s\sim\frac{1}{N_c}$ and should be not observable in the holographic regime. In spite of this, convincing arguments have been given that point to the fact that a form of screening is at work and affects observables, see \cite{Bigazzi:2008gd, Bigazzi:2008ie, Bigazzi:2008qq, Bigazzi:2009gu}. 

Aside from this, an (unwelcome) feature is that the addition of many flavours in the models above mentioned comes together with a badly defined UV QFT (in the sense that the UV is not field theoretical). This makes difficult the application of holographic renormalisation techniques \cite{Papadimitriou:2004ap}.

In this work we propose a pair QFT/holographic background that addresses the above problems. The model studied here belongs to the class of systems that use solutions of the Anabal\'on-Ross type \cite{Anabalon:2021tua,Anabalon:2022aig, Anabalon:2024che, Anabalon:2024qhf}, extended to holographic duals to confining QFTs and linear quivers with gapped IR in \cite{Nunez:2023nnl, Nunez:2023xgl, Fatemiabhari:2024aua, Chatzis:2024top, Chatzis:2024kdu,Barbosa:2024smw, Kumar:2024pcz}. In this way, the above mentioned papers constructed infinite families of backgrounds with a dual field theoretical high energy behaviour (reflected by the asymptotic AdS$_5$) and a well behaved low energy (reflected by a smooth geometry), together with the presence of backreacted {\it localised} sources (all preserving four supercharges). We use this to calculate Wilson loop VEVs for a given gauge group in a QFT described by a linear quiver. We observe the effects of the localised sources on the Wilson loop and study this for different quivers.
The most salient point we observe is that the probe F1 string dives into the bulk, not only along the usual radial/energy coordinate (here denoted by $r$), but also along the `linear quiver' coordinate, referred here as $z$. The energy of this probe changes as we elongate the two ends of it. It displays a conformal law for small separations, a confining law for intermediate ones and a screening law for long elongations. In this work we analyse this interesting phenomenology in detail.

Below, we describe the general idea of this work, give some details of the system we work with and present an outline of this work and its goals.

\subsection{General idea and outline of this work}\label{QFTapproach}
The main goal of this work is to study the cross-over between confinement and screening. By this, we mean the transition between an area law and a perimeter law for the Wilson loop VEV. For this to take place, matter transforming non-trivially under the center of the gauge group must take part in the dynamics. The typical example is QCD.

Holography is our tool of choice to study the problem. In the absence of a sharply defined holographic dual to QCD (see the discussion above), we choose to work with the well defined holographic dual to a family of 4-dimensional linear quivers with four supercharges and a strongly coupled conformal UV. These linear quivers are `balanced' hence the presence of flavours is immanent to the construction of the UV fixed point. The holographic set ups contain gauge groups with adjoint and bifundamental matter and also matter in the fundamental representation of some of the gauge groups. The `fundamental matter'
is holographically realised by the presence of localised D-brane sources in the bulk. These sources allow the possibility of open strings ending on them.

We calculate the rectangular Wilson loop VEV following 
the usual prescription \cite{Maldacena:1998im}, \cite{Rey:1998ik} (for a summary see \cite{Sonnenschein:1999if}, \cite{Nunez:2009da}). Indeed, hanging a fundamental string probe, from the asymptotic radial direction (the UV of the dual QFT) we separate the end-points of the probe, along the $x$ direction in the QFT. The string moves inside the bulk, minimizing its action. The F1 probe `falls' into smaller values of the radial $r$-coordinate, also exploring the `quiver tail' direction (denoted by $z$ in this paper).

The minimization of the action for this probe is a problem that we do not attempt to solve analytically. Instead, we use a numerical approach described in detail later in the paper.

We find that the string probe enters into the radial (AdS) direction in a way suggestive of confinement. At the same time and along the quiver direction $z$, the string dives towards the position of the closest flavour brane source. Once the string is close to these D-branes sources, one must consider the process in which the F1 probe can snap and attach itself to the sources. This process, typically suppressed by the three-strings coupling, $g_s\sim\frac{1}{N_c}$ is enhanced by the presence of many localised sources, leading to $g_s N_f\sim\frac{N_f}{N_c}\sim 1$.

We plot the expressions for the energy of the probe $E$ in terms of the separation in the QFT $L$ (in the $x$ direction). We give an analytic expression for $E(L)$, that very well interpolates the numerical results. We also present plots for the string profiles, showing how these explore the radial coordinate ($r$) and the quiver direction ($z$).

As a proof of concept\footnote{The techniques developed here can be applied to other holographic duals with similar characteristics, like those in \cite{Fatemiabhari:2024aua}, \cite{Chatzis:2024top}, \cite{Chatzis:2024kdu}.}, we study a particular family of holographic backgrounds dual to four dimensional ${\cal N}=1$ SCFTs, that after breaking of conformality by a VEV, represents holographically the (SUSY preserving) compactification of the CFT$_4$. This family of backgrounds is probed by fundamental strings that calculate the rectangular Wilson loop VEV for the resulting gapped (2+1) QFT.

The material is organised as follows. Section \ref{section-geometry} is written as a three-steps procedure. In there, we carefully describe the construction of the family of backgrounds we work with. In Section \ref{section-wilson}, we write the generic action for the Wilson loop exploring both the radial-holographic direction and the quiver-tail $z$ direction (as we stretch the quark-antiquark separation in $x$). We write the equations of motion and make some qualitative analysis of the dynamics. In Section \ref{sec:numerical}, we describe the numerical technique and approach and carefully analyse the results. The interpretation of these results suggests a confinement-screening transition. Section \ref{concl} gives some conclusions and presents topics for future study. Appendix \ref{sec:RobinHood} gives an account of the program and how to use it, which may be beneficial for colleagues wishing to use the publicly available code.

\section{The Supergravity Background and the dual QFT}\label{section-geometry}
In this section we present the supergravity background that we use to compute Wilson loops VEVs. This section is written as a three-stages procedure.
\begin{enumerate}
 \item{We first present a family of supergravity backgrounds dual to  six-dimensional ${\cal N}=(1,0)$ conformal field theories of the linear quiver type. }
 \item{Then, we compactify the SCFT$_6$ on a hyperbolic manifold, flowing at low energies to a family of four dimensional ${\cal N}=1$ SCFTs. We write the supergravity configurations capturing the dual description of this RG-flow.}
 \item{Finally, starting from the fixed point associated with the SCFT$_4$, we perform a deformation driven by a VEV. This deformation leads to a family of smooth backgrounds, dual to gapped field theories in (2+1) dimensions.}
\end{enumerate}
Below, we detail the steps of this construction. To close the section, we discuss briefly the QFT at the end of this RG-flow.

\subsection{Step 1: Holographic dual to a family of SCFT\textitle{$_6$}}
Let us start with a brief summary of the massive Type IIA backgrounds dual to six dimensional ${\cal N}=(1,0)$ SCFTs. The family of solutions was written in \cite{Apruzzi:2013yva, Apruzzi:2015wna}. In this work we use the notation of \cite{ct2015}. The massive IIA backgrounds can be written in terms of a metric, NS two form $B_2$, dilaton $\Psi$ and Ramond fields $F_0$, $F_2$. They have $SO(2,6)\times SU(2)_R$ isometries, that realise the (bosonic) global symmetries of 6d SCFTs with eight Poincare supercharges (that the backgrounds also posses). These configurations read,
 \begin{eqnarray}
& & ds^2=f_1(z) ds^2_{AdS_7}+f_2(z) dz^2+f_3(z) d \Omega^2(\theta_2, \phi_2),\nonumber\\
& & B_2=f_4(z) \mathrm{Vol}{(S^2)},\;\;\; F_2= f_5(z) \mathrm{Vol}{(S^2)}, \;\;\; e^{\Psi}=f_6(z).\label{backgroundads7xm3}\\
& & f_1(z)= 8 \sqrt{2} \pi \sqrt{-\frac{\alpha}{{\alpha''}}},\;\;\; f_2(z)= \sqrt{2} \pi \sqrt{-\frac{{\alpha''}}{{\alpha}}},\;\; 
f_3(z)=\sqrt{2} \pi \sqrt{-\frac{{\alpha''}}{\alpha}}\left( \frac{\alpha^2}{{\alpha'}^2-2 \alpha {\alpha''}}\right),\nonumber\\
& & f_4(z)=\pi \left(-z +\frac{\alpha {\alpha'}}{{{\alpha'}}^2-2 \alpha {\alpha''}}\right),\;\;\; f_5(z)=\left( \frac{{\alpha''}}{162 \pi^2}+ \frac{\pi F_0 \alpha {\alpha'}}{ {\alpha'}^2-2 \alpha {\alpha''}} \right),\nonumber\\
& &f_6(z)=2^{\frac{5}{4}} \pi^{\frac{5}{2}}3^4 \frac{(-\alpha/ {\alpha''})^{\frac{3}{4}}}{\sqrt{{\alpha'}^2-2 \alpha {\alpha''}}}.
\label{functionsf}
\end{eqnarray}
We have defined $d \Omega^2(\theta_2, \phi_2)=d \theta_2^2 + \sin^2 \theta_2 ~d \phi_2^2$ and $\mathrm{Vol}{(S^2)}=\sin\theta_2\;d\theta_2\wedge d\phi_2$. 
The different geometries specified by the function $\alpha(z)$ are supersymmetric solutions of the Massive IIA equations of motion (with mass parameter $F_0$), if $\alpha(z)$ solves the differential equation
\begin{equation}
{\alpha'''}=-162 \pi^3 F_0.
\label{alphathird}
\end{equation}
Since $F_0$ is a constant piece-wise continuous function, $\alpha(z)$ must be a continuous cubic function of the form
\begin{equation}\label{alpha}
 \alpha(z)= a_0+ a_1 z +\frac{a_2}{2}z^2 -\frac{162\pi^3 F_0}{6} z^3.
\end{equation}
Given a six-dimensional ${\cal N}=(1,0)$ super-conformal field theory encoded in a quiver diagram, Cremonesi and Tomasiello \cite{ct2015} gave a recipe to find the precise solution to eq.(\ref{alphathird}), such that when replaced in eq.(\ref{backgroundads7xm3})-(\ref{functionsf})
gives the holographic dual to the UV-SCFT$_6$. The recipe is the following:
\begin{itemize}
 \item{Consider a linear quiver with $(P-1)$ gauge nodes. The quiver must be `balanced' to avoid gauge anomalies. We write the associated rank function $R(z)$,
 $$
R(z)=-\frac{1}{81\pi^2}\alpha''(z) = \left\{
 \begin{array}{ll}
N_1 z & \quad 0 \leq z\leq 1 \\
N_1 +(N_2- N_1) (z-1) & \quad 1\leq z\leq 2\\
N_k + (N_{k+1}-N_k)(z-k) & \quad k\leq z\leq (k+1)\\
.... \\
N_{P-1}(P-z) & \quad (P-1)\leq z\leq P. %
 \end{array}
 \right.
$$}
\item{
The ranks of the color groups are encoded in the values of the rank function at the integers $z=(1,2,3,4,....P-1)$. Also, the second derivative of the rank function encodes the rank of the $SU(F_k)$ gauge groups. In fact,
\begin{equation}
R''(z)=\sum_{k=1}^{P-1} F_k\delta(z-k),~~~~ \text{with}~~~ F_k=2N_k -N_{k-1}- N_{k+1},\label{balance}
\end{equation}
which is the balancing condition above mentioned.}
\item{The function $\alpha(z)$ can be found by integrating twice $-81\pi^2 R(z)$. The integration constants are determined by imposing continuity of $\alpha(z)$ and $\alpha'(z)$ in each interval. One also imposes $\alpha(0)=\alpha(P)=0$.}
\end{itemize}
When the function $\alpha(z)$ above is used in eqs.(\ref{backgroundads7xm3})-(\ref{functionsf}), one obtains the holographic dual to the six dimensional ${\cal N}=(1,0)$ quiver field theory of Figure \ref{fig:quiver} at the origin of the tensor branch.
\begin{figure}
\begin{center}
	\begin{tikzpicture}
	\node (1) at (-4,0) [circle,draw,thick,minimum size=1.4cm] {N$_1$};
	\node (2) at (-2,0) [circle,draw,thick,minimum size=1.4cm] {N$_2$};
	\node (3) at (0,0) {$\dots$};
	\node (4) at (2,0) [circle,draw,thick,minimum size=1.4cm] {N$_{P-1}$};
	\draw[thick] (1) -- (2) -- (3) -- (4) 
 ;
	\node (1b) at (-4,-2) [rectangle,draw,thick,minimum size=1.2cm] {F$_1$};
	\node (2b) at (-2,-2) [rectangle,draw,thick,minimum size=1.2cm] {F$_2$};
	\node (3b) at (0,0) {$\dots$};
 ;
	\node (4b) at (2,-2) [rectangle,draw,thick,minimum size=1.2cm] {F$_{P-1}$};
	\draw[thick] (1) -- (1b);
	\draw[thick] (2) -- (2b);
	\draw[thick] (4) -- (4b);
	\end{tikzpicture}
\end{center}
 \caption{A linear quiver. The balancing condition implies $F_k = 2 N_k - N_{k-1}-N_{k+1}$ for each node.}
 \label{fig:quiver}
\end{figure}
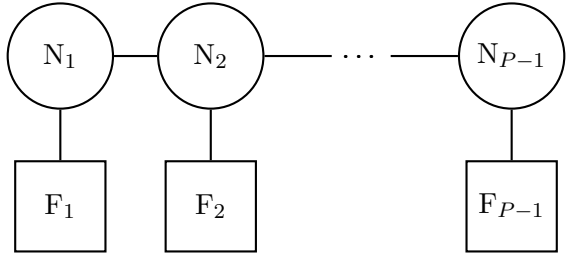
There are various checks of this duality. For example, Page charges have been computed for the Ramond and Neveu-Schwarz fields and put in correspondence with the quiver of Figure \ref{fig:quiver} and the associated Hanany-Witten set-up. Free energy of the SCFT$_6$ can be calculated holographically and compared with a field theory expression obtained in \cite{Nunez:2023nnl}. See \cite{Nunez:2018ags, Filippas:2019puw} for a detailed account of observables calculated holographically. 

We now move to the second step of the procedure: using the backgrounds in eq.(\ref{backgroundads7xm3}), to construct a family of AdS$_5$ solutions dual to 4d ${\cal N}=1$ SCFTs. 
\subsection{Step 2: Holographic dual to the flow SCFT\textitle{$_6\to$} SCFT\textitle{$_4$}}
Based on the family of backgrounds in eq.(\ref{backgroundads7xm3}), one can construct a flow from the six dimensional CFT to a four dimensional CFT, represented by a background with an AdS$_5$ factor. These flows were constructed in \cite{Merrikin:2022yho}, see also \cite{Apruzzi:2015zna, bpt2017} for the study of the fixed point AdS$_5$ solution. The family of backgrounds dual to 4d ${\cal N}=1$ SCFTs is obtained when we compactify, on a hyperbolic space H$_2$, the backgrounds in (\ref{backgroundads7xm3}). 

The backgrounds describing the flow between AdS$_7$ and AdS$_5$ are written in terms
of coordinates, parameters and functions,
\begin{eqnarray}
& & \text{Coordinates:}~(t,x_1,x_2,x_3, r, \theta_1,\phi_1, z, \theta_2,\phi_2).~~\text{Parameter:}~~e^{4\Psi_0}= 2^5 \times 81^4\times \pi^{10}.\label{funciones}\\
& & 
\text{Functions:}~\alpha(z), f(r), h(r), X(r)
, ~~\omega(r,z)= \left( \frac{\alpha'(z)^2 - 2 \alpha(z)\alpha''(z) X(r)^5 }{\alpha'(z)^2 -2 \alpha(z)\alpha''(z) } \right) .\nonumber
\end{eqnarray}
 The string-frame metric reads
 \begin{equation}\label{metric-z}
 \begin{aligned}
 ds^{2}_{st} &= 2\pi\sqrt{2}\sqrt{-\frac{\alpha(z)}{\alpha''(z)}}\,
 X(r)^{-\frac{5}{2}}
 \left[e^{2f(r)}dx^{2}_{1,3}
 + dr^{2} + e^{2h(r)}\left( d\theta^{2}_{1} + \sinh^{2}\theta_{1} d\phi^{2}_{1}\right)\right]\\
 & + X(r)^{5/2}\left[ \pi\sqrt{2}\sqrt{-\frac{\alpha''(z)}{\alpha(z)}}dz^{2} 
 +\frac{\sqrt{2}\,\pi}{\omega(r,z)} \frac{\sqrt{-\alpha^{3}(z)\alpha''(z)}}{2\alpha(z)\alpha''(z)-\alpha'^{2}} \left(d\theta^{2}_{2}+\sin^{2}\theta_{2}
 \left( d\phi_{2} +\cosh\theta_{1} d\phi_{1}\right)^{2}\right)\right].
 \end{aligned}
 \end{equation}

 %
The Neveu-Schwarz ($B_2,\Psi$) and Ramond ($F_0, F_2, F_4$) fields are,
 \begin{eqnarray}\label{NSRRforms}
 & & B_{2} = \left(\frac{\pi}{\omega(r,z)}\frac{\alpha(z)\alpha'(z)}{\alpha'(z)^{2}-2\alpha(z)\alpha''(z)}\sin \theta_{2} d\theta_{2} -\pi\cos \theta_{2} dz \right)\wedge\left( d\phi_{2} +\cosh \theta_{1} d\phi_{1} \right),\nonumber\\
 & & e^{4\Psi(r,z) }= \frac{X^5(r)}{\omega^2(r,z)} \left( \frac{-\alpha(z)}{\alpha''(z)}\right)^3\left( \frac{e^{2\Psi_0}}{ \alpha'(z)^{2}-2\alpha(z)\alpha''(z) }\right)^2,\nonumber\\
 & & F_0= -2^{\frac{1}{4}} \frac{e^{-\Psi_{0}}}{\sqrt{\pi}} \alpha'''(z) ,\label{eqF0}\\
 & & F_{2} = 2^{\frac{1}{4}}\sqrt{\pi}e^{-\Psi_{0}}\alpha''(z) \left[ \cos \theta_{2}\Vol(\Sigma_{2}) -\Vol(S^{2}_{c}) \right] 
 + F_{0}\frac{\pi}{\omega(r,z)} \frac{\alpha(z)\alpha'(z)}{\alpha'(z)^{2}-2\alpha(z)\alpha''(z)}\, \Vol(S^{2}_{c}),\nonumber\\
& & F_{4} = \left(\frac{2^{\frac{1}{4} } \pi^{\frac{3}{2} } e^{-\Psi_{0}} }{\omega(r,z)} \right) \left(\frac{\alpha(z)\alpha'(z)\alpha''(z)}{\alpha'(z)^{2}-2\alpha(z)\alpha''(z)}\right) \cos \theta_{2}\, \Vol(\Sigma_{2})\wedge \Vol(S^{2}) \nonumber\\
& &~~~~~+ 2^{\frac{1}{4}}\pi^{\frac{3}{2}}e^{-\Psi_{0}}\alpha''(z) \sin^{2}\theta_{2}\, dz\wedge d\phi_{2}\wedge \Vol(\Sigma_{2}).\nonumber
\end{eqnarray}
We have defined the volume elements,
\begin{eqnarray}
& & \Vol(S^{2}) = \sin\theta_{2} d\theta_{2} \wedge d\phi_{2},~~
 \Vol(S^{2}_{c}) = \sin(\theta_{2})d\theta_{2} \wedge\left( d\phi_{2}+\cosh \theta_{1} d\phi_{1}\right),\nonumber\\
 & & \Vol(\Sigma_2)= \sinh \theta_1 d\theta_1\wedge d\phi_1.\label{volumeforms} %
 \end{eqnarray}
It was shown in \cite{Merrikin:2022yho} that the configuration solves all the massive IIA equations of motion once eq.(\ref{alphathird}) and BPS equations for $f(r), h(r), X(r)$ (see equation.(2.5) in \cite{Merrikin:2022yho}), are imposed. The Bianchi identities for $F_0$ are violated, $dF_0=R''(z) dz$ at the position of the localised D8 brane stacks, as indicated by eq.(\ref{balance}).

In the UV of the system, for $r\to\infty$, the BPS equations of \cite{Merrikin:2022yho} have the solution,
\begin{equation}
 e^{2f(r)}\sim e^{2h(r)}\sim e^{2r},~~~X(r)=1. 
\end{equation}
The configuration in eqs.(\ref{metric-z})-(\ref{volumeforms}) approaches (for $r\to\infty$) that of 
eq.(\ref{backgroundads7xm3}), up to a gauge transformation on the NS two form (es explained in \cite{Merrikin:2022yho}) and the fact that AdS$_7$ is written as a foliation over $R^{1,3}\times H_2(\theta_1,\phi_1)$.

For low energies, the fixed point solution, attained at the end of the flow ($r\to -\infty$) is,
\begin{equation}
e^{2f(r)}= e^{\frac{2}{3} r},\;\;\; e^{2h(r)}=\frac{3}{4},\;\;\; X^5(r)= \frac{3}{4}.\label{fixedpoint}
\end{equation}
The fixed point solution is dual to a family of non-Lagrangian 4d ${\cal N}=1$ SCFT. The charges of D8, D6, NS5 are calculated in \cite{Merrikin:2022yho} and shown to be quantised. These numbers appear in the free energy and other observables of the 4d SCFT, as shown in Appendix B of \cite{Chatzis:2024kdu}.
The family of 4d ${\cal N}=1$ SCFTs has (bosonic) global symmetries $SO(2,4)$ associated with the isometries of AdS$_5$, $U(1)_R$ associated with isometries of the internal space and $\prod_{k=1}^{P-1} SU(N_k)$ ``flavour symmetry'' realised by the fields (vectors, scalars and fermions) describing the dynamics of the D8 sources. 

Let us now describe the third and last step in our construction.
\subsection{Step 3: Holographic dual to the deformation SCFT\textitle{$_4 \to$} QFT\textitle{$_3$}}
Our last step is to break conformality and flow to a gapped theory. To do this we follow the technique explained in \cite{Chatzis:2024top, Chatzis:2024kdu}. This is, a twisted compactification on a circle, preserving four supercharges. After this procedure is followed, we encounter a family of backgrounds, parametrised as above, by the function $\alpha(z)$.
The metric and dilaton read,
\begin{eqnarray}
 & & \mathrm{d}s_{10}^2=18\pi\sqrt{-\frac{\alpha}{6 {\alpha''}}}\left[\mathrm{d}s_5^2+\frac{1}{3} \mathrm{d}s_\Sigma^2-\frac{{\alpha''}}{6\alpha}\mathrm{d}z^2-\frac{\alpha {\alpha''}}{6 {\alpha'}^2-9 \alpha {\alpha''}} \left(\mathrm{d}\theta_2^2+\sin^2 \theta_2 \mathcal{D}\phi_2^2\right)
 \right],\nonumber\\
 & &e^{-4\Psi}= \frac{1}{2^5 3^{17}\pi^{10}}\left( -\frac{{\alpha''}}{\alpha}\right)^3 \left( 2{\alpha'}^2-3 \alpha {\alpha''}\right)^2,~~~{\cal D}\phi_2= \mathrm{d}\phi_2 - 3\mathcal{A} + A_\Sigma \label{metric-dil-BPT}\\
 & & \mathcal{A}=q\left(\frac{1}{r^2}-\frac{1}{r_*^2} \right) \mathrm{d}\phi,~~A_\Sigma= \cosh \theta_1 d\phi_1,
 ~~~ \mathrm{d}s_\Sigma^2= d\theta_1^2 + \sinh^2\theta_1 d\phi_1^2.
 \nonumber\\
 & &\mathrm{d}s^2_{5}=\frac{r^2}{l^2} (-\mathrm{d}t^2+\mathrm{d}x_1^2 + \mathrm{d}x_2^2 + f(r)\mathrm{d}\phi^2) + \frac{l^2 ~\mathrm{d}r^2}{ r^2 f(r)},~~~~ f(r) = 1 -\frac{\mu}{r^4}- \frac{q^2l^2}{r^6}.\label{ds^2AR}
\end{eqnarray}
The Ramond and Neveu-Schwarz potentials and their associated field strengths can be compactly written as
\begin{eqnarray}
& & B_2 =\frac{1}{3} \xi \wedge \mathcal{D}\phi_2, 
 \qquad 
 C_1 = \frac{{\alpha''}}{162\pi^2}\cos \theta_2\, \mathcal{D}\phi_2, \qquad 
 C_3 = \frac{{\alpha'}}{162\pi} \mathcal{D}\phi_2 \wedge \text{vol}_\Sigma, \nonumber\\
& &
H_3=\mathrm{d}B_2, 
 \qquad 
 F_2 =F_0 B_2 + \mathrm{d}C_1,\label{RR-NS-BPT}
\\
& & F_4 =\left(\mathrm{d}C_3 +B_2 \wedge F_2 -\frac{1}{2} F_0\, B_2 \wedge B_2\right) -\frac{{\alpha''}}{18\pi}\mathrm{d}z \wedge \left(\star_5 \mathcal{F}-\frac{1}{3}\mathcal{F} \wedge \mathcal{D} \phi_2\right) -\frac{{\alpha'}}{54\pi}\mathcal{F}\wedge \text{vol}_\Sigma , \nonumber
\end{eqnarray}
in terms of a one form $\xi$, the two form $\mathcal{F}=\mathrm{d}\mathcal{A}$, and its Hodge dual in five dimensions,
\begin{equation}
 \xi=3\pi \left( \cos\theta_2 \mathrm{d}z-\frac{2\alpha \dot{\alpha}}{2\dot{\alpha}^2-3\alpha \Ddot{\alpha}} \sin \theta_2 \mathrm{d}\theta_2 \right),~~\mathcal{F}= -\frac{2 q}{r^3} \mathrm{d}r\wedge \mathrm{d}\phi,~~\star_5 \mathcal{F}=-2q~ \mathrm{d}t\wedge \mathrm{d}x_1\wedge \mathrm{d}x_2.\label{fluxesqft3}
\end{equation}
We have checked that all equations of motion (Einstein, Maxwell and Bianchi)
are solved once eq.(\ref{alphathird}) is imposed. The Bianchi identity $dF_0= R''(z) dz$ indicate the presence of sources, as follows from eqs. (\ref{alphathird}) and (\ref{balance}). When the parameter $\mu=0$ in the function $f(r)$ of eq.(\ref{ds^2AR}), the background preserves four supercharges, otherwise, SUSY is broken.

Let us briefly comment on the QFT aspects of the background in eqs.(\ref{metric-dil-BPT})-(\ref{fluxesqft3}). We have the dual to a family of QFTs that describe the flow from a four dimensional CFT to a (2+1)-dimensional QFT. As discussed in \cite{Chatzis:2024top},\cite{Chatzis:2024kdu}, the background describes a twisted compactification on the $\phi$-circle. If the parameter $\mu=0$, the compactification preserves SUSY, thanks to the fibration represented by the one form ${\cal A}$. A lagrangian version of the QFT (for the case of compactifying ${\cal N}=4$ SYM) was discussed in \cite{Kumar:2024pcz}. A similar mechanism should take place for our non-lagrangian field theories. An interesting universality of observables, proposed in \cite{Gauntlett:2007ma}, is natural to infer using the holographic description--see \cite{Chatzis:2024top},\cite{Chatzis:2024kdu}. 
In this work, we are concerned with non-universal behaviours for different members of our backgrounds/field theories, namely on observables that do depend on the particular function $\alpha(z)$. In particular, we focus on the Wilson loop. The Wilson loop calculated for a given gauge node should explore both the radial direction $r$ and the $z$ direction labelling the quiver. Whilst the behaviour of a probe string exploring only the $r$ direction is universal, the ability of the probe to explore the $z$ direction brings the dependence on the rank function and the function $\alpha(z)$, as we discuss below. In particular, when the string probe comes closer to a stack of D8 (flavour) branes, the mechanism of screening takes place, due to the competition between a connected and a disconnected strings configuration. We move on to study this.

\section{Wilson Loops}\label{section-wilson}

 In this section, we calculate the Wilson loop expectation values for our family of QFTs whose dual description is in terms of the backgrounds of eqs.(\ref{metric-dil-BPT})-(\ref{fluxesqft3}). We intend to test the cross-over between confinement and screening behaviour at large separations for the external quarks in the QFT. We begin with the general formalism in holography, which is used to calculate the Wilson loop expectation values in QFTs, following the methods of \cite{Sonnenschein:1999if, Nunez:2009da}. 
 The procedure includes embedding a probe fundamental string into the gravity dual.
 We will study the probe related to the Wilson loop in three stages, considering the embeddings in different submanifolds of the geometry.
\subsection{Case I: x, t, r[x]}\label{sec:rofx}
For a generic background with the metric 
 \begin{equation}
 ds^{2}=-g_{tt}dt^{2}+g_{xx}d\vec{x}^{2}+g_{rr }
 dr^{2}+g_{ij}d\theta ^{i}d\theta ^{j}\, . \label{wilsonb}
 \end{equation} 
One can assume an embedding for a probe string with a Nambu-Goto action as 
\begin{eqnarray}
& & t=\tau,~~~x=x(\sigma),~~~r=r(\sigma).\nonumber\\
& & S_{NG}= T_{F1} \int d\tau d\sigma \sqrt{g_{tt}(r) g_{xx}(r)x'^2 + g_{tt}(r)g_{rr}(r) r'^2}.\label{NGwilson}
\end{eqnarray}
Here, and for all subsequent sections, $x=x_1$. All other coordinates in the background are kept fixed. The coordinates $(\tau,\sigma)$ parameterise the string worldsheet.
The equations of motion for the string read, see \cite{Nunez:2009da},
 \begin{equation}
 \frac{dr }{d\sigma }=\pm \frac{dx}{d\sigma }V_{eff}\left( r \right) \, .\label{eqmov}
 \end{equation}
Following \cite{Nunez:2009da}, we define the `effective potential' as,
 \begin{equation}
 V_{eff}\left( r \right) =\frac{F\left( r \right) }{CG\left( r
 \right) }\sqrt{F^{2}\left( r \right) -C^{2}}\ ,~~ F^2\left( r \right) =g_{tt}g_{xx},~~G^2\left( r \right)=g_{tt}g_{rr},\label{potdef}
 \end{equation}
 where the constant $C=\frac{F^2x'}{\sqrt{F^2 x'^2+ G^2 r'^2}}$ is calculable using the equations of motion. 
 Indeed, by choosing $x(\sigma)=\sigma$, and since the action does not explicitly depend on $\sigma$, one finds a conserved `Hamiltonian'. Using eq.(\ref{eqmov}) we have $C= F(r_0)$, where $r_0$ is the point where the embedded string turns in the U-shape embedding, satisfying $r'(\sigma)=0$. We choose $C=F(r_0)$ for this subsection and hence fix this freedom.

In this framework, we have an open string with its endpoints attached to a D-brane located at $ r \to \infty $. Dirichlet boundary conditions are applied to the string at $ r \to \infty $ by ensuring that $ V_{eff}|_{r \to \infty} $ approaches infinity. The distance between the string's endpoints can be interpreted as the separation between a quark and an antiquark (both non-dynamical) in the dual field theory. The energy of this quark-antiquark pair is derived from the Nambu-Goto action. We need to regularize this energy, subtracting the energy of two static strings extending over the entire radial range $[r^*, \infty)$, which accounts for the rest mass of the quark-antiquark pair.

As stated, the string adopts a U-shape in the bulk. The separation and energy of the quark-antiquark pair can be expressed in terms of the distance from the turning point of the string, \( r_0 \), as follows:
 \begin{align}
 L_{QQ}\left( r_{0}\right) &=2\int_{r_{0}}^{+\infty }
 \frac{d\r}{V_{eff}(\r) }\, , \label{QQ separation} \\
 E_{QQ}\left( r_{0}\right) &=F\left( r_{0}\right) L_{QQ}\left( r_{0}\right)
 +2\int_{r_{0}}^{+\infty }d\r\frac{G\left( \r\right) }{F\left( \r\right) }
 \sqrt{F\left( \r\right) ^{2}-F\left( r_{0}\right) ^{2}}-2\int_{r^*}^{+\infty }d\r\
 G\left( \r\right) \label{QQ energy} \, .
 \end{align}

In \cite{Nunez:2009da} the criteria for confinement/screening are given.
The mentioned general procedure can be applied to the background discussed in eq.(\ref{metric-dil-BPT}). Note that we have kept $z$ fixed.
In fact, we assume that the string is located at a constant integer value $z=z^*$ (this implies that we are calculating the Wilson loop in the $z^*$-th gauge group). The embedding coordinates are chosen as $t=\tau,\, x=\sigma, \,r=r(\sigma)$. We have
\begin{equation}
S_{NG}= T_{F1}\int d\tau d\sigma \sqrt{\det[g_{\alpha \beta}]} =T_{F1} T\int_{-L/2}^{+L/2} \dd x \sqrt{F^2+G^2 r'^2}. \label{NGwilson2}
\end{equation}
with function definitions
\begin{equation}
F^2=-\frac{\alpha(z)}{\ddot\alpha(z)} r^4,~~ ~~~~
G^2=-\frac{\alpha(z)}{\ddot\alpha(z)}\frac{1}{f(r)},~~ ~~~~
\end{equation}
In this section we denote $\alpha'(z)=\frac{d\alpha(z)}{dz} \equiv \dot{\alpha}(z)$ and similarly for $\alpha''(z)=\frac{d\alpha(z)}{dz^2} \equiv \ddot{\alpha}(z)$, reminding that in this discussion $z$ is fixed at $z^*$. For the effective potential one has
\begin{equation}
V_{eff}=\frac{\sqrt{f(r)}r^2}{r_0^2} \sqrt{r^4-r_0^4}.
\end{equation}

We investigate the length and the energy of the quark-antiquark pair for analysing the behaviour of the QFT at low energies and searching for confinement (or screening) behaviour.
Using eqs.(\ref{QQ separation})-(\ref{QQ energy}), one has the length and energy of the quark-antiquark pair as
\begin{eqnarray}
& & L_{QQ}\left( r_{0}\right) = 2 \int_{r_{0}}^{\infty }\frac{r_0^2}{r^2 \sqrt{f(r)(r^4-r_0^4)}} dr\ ,\label{Lqq} \\
& & E_{QQ} \left( r_{0}\right) =F(r_{0}) L_{QQ} \left( r_{0}\right) +
{2}\sqrt{-\frac{\alpha(z^*)}{\ddot\alpha(z^*)}} \int_{r_{0} }^{\infty } dr \frac{\sqrt{r^4-r_0^4}}{\sqrt{f(r)}r^2} +\nonumber\\
 & & 
\qquad\qquad\qquad -{2}\sqrt{-\frac{\alpha(z^*)}{\ddot\alpha(z^*)}} \int_{r^* }^{\infty } dr \sqrt{\frac{1}{f(r)}}
 \ .\label{Eqq}
\end{eqnarray}%
The integrals can be calculated using numerical methods. The criteria for the stability of the embeddings are discussed in detail in \cite{Arias:2009me,Chatzis:2024dlt} and can be easily implemented for the present case.

 \subsection{Case II: x, t, z[x]}
In our dual QFT setups, massless flavour quarks are present. Specifically, for each kink in the convex-piecewise linear rank function \({\cal R}(z)\), there is a corresponding flavour group, which consists of a set of D-branes localized in the $z$ direction.

These flavour groups enable the screening phenomenon, which can occur through the creation of a pair of flavour quarks that disrupt the flux tube connecting the heavy probe quark to the anti-quark pair.

\begin{figure}
\begin{center}
	\begin{tikzpicture}
	\node (1) at (-4,0) [circle,draw,thick,minimum size=1.4cm] {N$_1$};
	\node (2) at (-2,0) [circle,draw,thick,minimum size=1.4cm] {N$_2$};
	\node (3) at (0,0) {$\dots$};
	\node (5) at (4,0) [circle,draw,thick,minimum size=1.4cm] {N$_{P}$};
	\node (4) at (2,0) [circle,draw,thick,minimum size=1.4cm] {N$_{P-1}$};
	\draw[thick] (1) -- (2) -- (3) -- (4) -- (5);
	\node (3b) at (0,0) {$\dots$};
	\node (4b) at (2,-2) [rectangle,draw,thick,minimum size=1.2cm] {F$_{P-1}$};
	\draw[thick] (4) -- (4b);
 \draw[thick,red] (-4,-0.82) -- (1.7,-0.82);
 \draw[thick,red] (1.7,-0.82) -- (1.7,-1.4);
 \draw[thick,red] (1.6,-1.4) -- (1.6,-0.92);
 \draw[thick,red] (1.6,-0.92) -- (-4,-0.92);
 \node at (-4,-0.82) {\textbullet};
 \node at (-4,-0.92) {\textbullet};
	\end{tikzpicture}
\end{center}
 \caption{Schematic example plot for the insertion of a probe quark anti-quark pair at the node $N_1$ in a linear quiver. Screening becomes possible if the chain of interactions through the field theory degrees of freedom can excite a pair from flavour fermions in the $F_{P-1}$ group.}
 \label{fig:screeningquiver}
\end{figure}
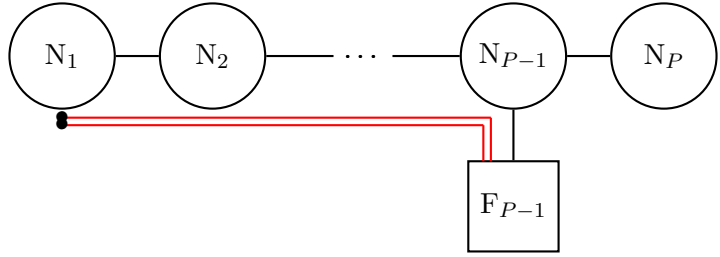

Even if the Wilson loop is associated with a gauge node that lacks flavour groups, interactions across the quiver can still excite the flavour quarks, as pictured in Fig. \ref{fig:screeningquiver}.
{The operator mediating this interaction is of the form 
\begin{equation}
O=Q^{a_1} B^{a_1 a_2} B^{a_2 a_3}.... B^{a_{P-2} b_{P-1} } q^{a{P-1}}\bar{q}^{a_{P-1}} \bar{B}^{a_{P-1} a_{P-2}}....\bar{B}^{a_2 a_1} \bar{Q}^{a_1}.\label{operatorred}
\end{equation}
We have denoted by $(Q,\bar{Q})$ the external quarks, by $(q,\bar{q})$ the dynamical quarks and with $(B,\bar{B})$ the bifundamentals connecting gauge nodes (a similar operator, in a different system, was studied in \cite{Bergman:2018hin}). Clearly, the longer is the distance (in field space) between the gauge node and the nearest flavour group, the heavier is the operator connecting them, and the more suppressed the screening process is.}
To study this process in detail, it is necessary to insert a probe string into the bulk, allowing it to extend not only in the spatial $x$ and $r$ directions but also in the $z$ direction, which is related to the gauge node in the quiver. We will perform this calculation in two steps.

 In this section, we embed a string in the background in eq.(\ref{metric-dil-BPT}) that is extended in the $t,\,x$ and $z$ directions to further examine the screening behaviour of the dual QFTs. Following the same reasoning of the previous section, for a string embedding parameterised by $(\tau,\sigma)$ coordinates and extended in $t,\,x$ and $z$ directions, we choose
 \begin{equation}
 t=\tau,~~x=\sigma,~~z=z(x),~~ r=\bar{r}~~ (\text{fixed}).
 \end{equation} 
The Nambu-Goto action is
\begin{eqnarray}
& &S_{NG}= T T_{F1}\int d\sigma \sqrt{F^2 + S^2 z'^2},\label{NGgeneric2}\\
& & F^2=-\frac{\alpha(z)}{\ddot\alpha(z)} \bar{r}^4,~~ ~~~~ S^2=\frac{\bar{r}^2}{6}.\nonumber
\end{eqnarray}

To determine if a configuration extending in the \( z \) direction can approach the nearest flavour group, we need to minimize this action. Further generalization involves considering a configuration that can extend freely in both \( r \) and \( z \) directions, which we address in the next section.
For the present case, we have 
\begin{equation}
 S_{NG}= T_{F1} T ~\int~d z \sqrt{ -\frac{\alpha(z)}{\ddot\alpha(z)} \bar r^4+\frac{\bar r^2}{6}z'^2}.
\end{equation}
One can investigate this action for different rank functions. 
We assume that the quark-antiquark pair is introduced at the first gauge node, which corresponds to the boundary condition for the string at \( z^*=1 \). The string is expected to extend into the bulk in the \( z \) direction up to a point \( z_0 \) and then return, forming a U-shaped configuration. Based on the analysis of equations (\ref{QQ separation}) and (\ref{QQ energy}), we have:
\begin{align}
V_{eff}\left( z \right) &=\frac{F\left( z \right) }{F\left( z_{0}\right)S\left( z
 \right) }\sqrt{F^{2}\left( z \right) -F^{2}\left( z_{0}\right)}\ ,\\
 L_{QQ}\left( z_{0}\right) &=2\int_{z^*}^{z_{0} }
 \frac{dz}{V_{eff}(z) }\, , \label{QQseparationet} \\
 E_{QQ}\left( z_{0}\right) &=F\left( z_{0}\right) L_{QQ}\left( z_{0}\right)
 +2\int_{z^*}^{z_{0}}dz\frac{S\left( z\right) }{F\left( z\right) }
 \sqrt{F\left( z\right) ^{2}-F\left( z_{0}\right) ^{2}}
 \label{QQenergyet} \, .
 \end{align}

Let us now discuss the physically more relevant situation in which the string explores the $(r,z)$-space as is stretched in $x$.
 \subsection{Case III: x, t, r[x], z[x]} \label{sec:genericem}

In general, the embedded string in the directions $t,\,x,\,r,\,z$ can be again parameterised by world sheet coordinates $(\tau,\sigma)$, according to 
\begin{eqnarray}
& & t=\tau,~~x=x(\sigma),~~r=r(\sigma),~~z=z(\sigma).\label{parametri}
\end{eqnarray}
By integrating over $0\leq\tau\leq T$, the Nambu-Goto action for the probe string is,
\begin{eqnarray}
& &S_{NG} = T T_{F1} \int \dd \sigma \sqrt{F^2 x'^2+ G^2 r'^2+ S^2 z'^2},\label{NGgeneric}\\
& & 
F^2=-\frac{\alpha(z)}{\ddot\alpha(z)} r^4,~~ ~~~~
G^2=-\frac{\alpha(z)}{\ddot\alpha(z)}\frac{1}{f(r)},~~ ~~~~
S^2=\frac{r^2}{6},\nonumber
\end{eqnarray}
By choosing $x=x(\sigma)=\sigma$ the action simplifies as 
\begin{eqnarray}\label{NGaction}
& &S_{NG} = T T_{F1} \int \dd \sigma \sqrt{F^2 + G^2 r'^2+ S^2 z'^2},
\end{eqnarray}
with derivatives now respect to $x$.
To determine if a configuration extending in the $z$ and $r$ direction can approach the nearest flavour group, we need to minimize this action. In this case there is one conserved `Hamiltonian' but as the action depends explicitly on $z$, one cannot find the second integral of motion straightforwardly. 
Because the action does not depend on $x$ explicitly, one can find the conserved `Hamiltonian' as
\begin{equation}
 C \equiv \frac{-F^2}{\sqrt{F^2 + G^2 r'^2+ S^2 z'^2}}.
\end{equation}
If the string embedding takes a U-shaped form, one can derive the relation $C=-F(r_0,z_0)$, where $(r_0,z_0)$ are the coordinates of the turning point. The two equations of motion obtained from the action read
\begin{eqnarray}
& & \frac{d}{d x}\left( \frac{G^2 r'}{\sqrt{F^2 + G^2 r'^2+ S^2 z'^2}}\right) = \frac{F \partial_r F + G \partial_r G ~r'^2+S \partial_r S ~z'^2}{\sqrt{F^2 + G^2 r'^2+ S^2 z'^2}}~, \nonumber\\
& & \frac{d}{d x}\left( \frac{S^2 z'}{\sqrt{F^2 + G^2 r'^2+ S^2 z'^2}}\right) = \frac{F \partial_z F + G \partial_z G ~r'^2+S \partial_z S ~z'^2}{\sqrt{F^2 + G^2 r'^2+ S^2 z'^2}}~.\label{eqB3}
\end{eqnarray}

One may solve these equations using numerical methods. The relevant domain restrictions and boundary conditions for a U-shaped embedding ranging over $x \in [-L/2,L/2]$ is 
\begin{align}
 &r(x) \in [r^*,\infty],\quad r(x=\pm L/2)=\infty, \quad r(x=0) = r_0;\nonumber\\
 &z(x) \in [0,P],\quad z(x=\pm L/2)=z^*, \quad z(x=0) = z_0; \label{bcs}\\
 &r^* \in [0,\infty),\quad z^* \in [0,P],\qquad z^*,P\in\mathbb{N}. \nonumber
\end{align}
The differential equations are highly non-linear, coupled boundary value problems. In the process of numerically solving these equations, we encounter difficulties. In fact, imposing the boundary conditions, especially in the $z$ direction, makes it challenging to obtain the desired solutions. Hence, in what follows we approach this problem by studying the minimization of the action in eq.(\ref{NGaction}). The method used is described in the next section.

We find, for any fixed $L$ (the separation in $x$), the $r(x)$ and $z(x)$ that minimize the action, that is a solution of the equations of motion. The parameters that interest us are $(L, E)$, where the energy $E$ is simply defined as the action $S_{NG}$ evaluated on the solution found.

\FloatBarrier
\section{Numerical approach and results}\label{sec:numerical}

As mentioned in Section \ref{section-wilson}, solving the equations of motion \eqref{eqB3} is a daunting problem. Instead, we choose to find their solutions minimizing the action \eqref{NGaction} via a numerical optimization algorithm written in the Julia programming language, publicly available at \url{https://github.com/cu2mauro/RobinHood.jl}. Some more technical details about the algorithm, including how to use it, are presented in Appendix \ref{sec:RobinHood}. Here we show the procedure followed to find the mentioned solutions, presenting the outline of the algorithm, the solutions found for three different rank functions, and our analysis of these results.

\FloatBarrier
\subsection{Algorithm}\label{algor}

The functions $r(x),\,z(x)$ are approximated via the use of splines \cite{splines}, i.e. piecewise polynomial functions, and in particular linear 1D splines. The interval $[-L/2,L/2]$ is subdivided into $n$ sub-intervals: 

\be
[x_i,x_{i+1}],\quad i=0\dots n-1, \quad x_0=-L/2,\,x_n=L/2.
\ee
Both $r$ and $z$ are taken as arrays of length $n$, so that the $i$-th element of each can be used to approximate the value of the corresponding function at $x_i$. The action is a function of the two splines and the interval, which performs a simple trapezoidal quadrature, integrating the Lagrangian. Then the action in eq.(\ref{NGaction}), together with the boundary conditions \eqref{bcs}, is passed to an optimization solver, which uses an interior-point Newton method \cite{NoceWrig} to find which values for the arrays $r$ and $z$ minimize the action while respecting the domain restrictions and boundary conditions. To calculate the derivatives of the numerical function $S_{NG}(r,z)$ we employ automatic differentiation in reverse accumulation, which is the fastest option for our problems. 

The value of the energy $E$ is the action evaluated on the minimal configuration, and together with the separation $L$ and the minimal splines $r_m(x),\,z_m(x)$ it is stored to be analysed. 

Therefore, specifying the desired $\alpha(z)$ function, the parameters of the theory $(l,\mu,q,P,N,z^*)$, and the separation $L$, our algorithm can find the configuration that solves the corresponding equations of motion and find the energy of such configuration. Of all these variables, we explore how different $\alpha(z)$ behave when changing $(L,P,z^*)$, fixing $\mu=0$ (to be in the simpler SUSY case) and $q=1,\,l=1$. The parameter $N$ can be fixed to any large number, being only present as an overall factor in the cases of choice. We remind the reader that the parameter $z^*$ is an integer number in $[1, P-1]$ indicating the position (in the $z$-coordinate) of the gauge node for which we compute the Wilson loop.
\\
Below, we present the results in the case of three numerical experiments. We follow the same logic and order in the presentation of each of them, with emphasis on the physical meaning of our results.

\FloatBarrier
\subsection{Solutions}\label{sec:solutions}

Using the algorithm presented in \ref{algor}, we perform three different ``experiments''. We chose three different functions $\alpha(z)$ in eq.(\ref{alpha}) resulting in three different rank functions $R(z)$. For each experiment, we run the algorithm for multiple values of $(L,P,z^*)$ and record, for each iteration, the functions $r_m(x),\,z_m(x)$ that minimize the action and the value of the minimized action, i.e. the energy $E$. Here we present the data and the solutions found.

\FloatBarrier
\subsubsection{Experiment 1: scalene triangle rank}\label{rankscalene}

For the first experiment, we chose a quiver that features a flavour group at $z=P-1$, resulting in $\alpha(z)$ and a triangular rank function, as displayed in Figure \ref{fig:exp0alpha}.

\begin{figure}[!htbp]
\centering
 \includegraphics[width=.50\textwidth]{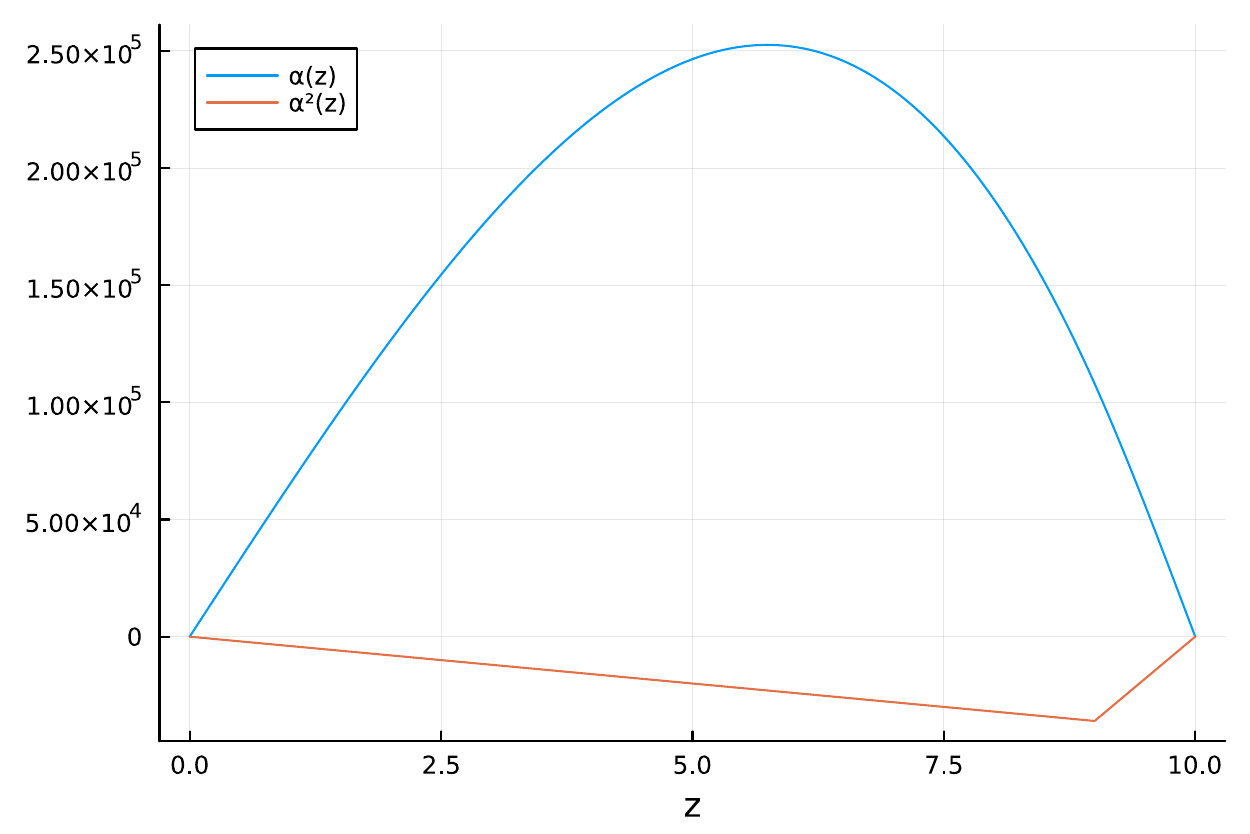}
\caption{The $\alpha(z)$ and $\alpha''(z)$ functions for the scalene triangular quiver. The rank function presents a kink at $z=P-1$, the position of the sources (flavour branes).\label{fig:exp0alpha}}
\end{figure}

In fact, for 
$$
R(z)=-\frac{1}{81\pi^2}\alpha''(z) = \left\{
 \begin{array}{ll}
N z & \quad 0 \leq z\leq (P-1), \\
N(P-1) (P-z) & \quad (P-1)\leq z\leq P;
 \end{array}
 \right.
$$
the corresponding quiver is
\begin{center}
	\begin{tikzpicture}
	\node (1) at (-6,0) [circle,draw,thick,minimum size=1.2cm] {N};
	\node (2) at (-4,0) [circle,draw,thick,minimum size=1.2cm] {2N};
	\node (3) at (-2,0) [circle,draw,thick,minimum size=1.2cm] {3N};	
	\node (4) at (0,0) {$\dots$};
	\node (6) at (4,0) [rectangle,draw,thick,minimum size=1.2cm] {PN};
	\node (5) at (2,0) {(P-1)N};
	\draw[thick] (1) -- (2) -- (3) -- (4) -- (5)-- (6);
	\draw[thick] (2,0) circle (0.7cm) ;
	\draw[thick] (1,0) -- (1.3,0);
	\draw[thick] (2.7,0) -- (3.3,0);
	\end{tikzpicture}\
\end{center}

We study this quiver to understand how the solutions for the probe string vary when changing the value of $P$. Also, this illustrative example gives us a structure to better read off results from the following quivers.

We set our Wilson loop in the first gauge node with $z^*=1$. Choosing $P$ larger makes it energetically more expensive to create a dynamical quark-antiquark pair. In fact, as $P$ grows large, the operator in eq.(\ref{operatorred}) -- see Figure \ref{fig:screeningquiver} -- connecting the QCD-string to the flavour group is heavier and more difficult to excite.
The results of this analysis are visible in Figures \ref{fig:exp0rz}, \ref{fig:exp0string} and \ref{fig:exp0LE}.
\\
To begin with, we set $P=10$ (that is a quiver with nine gauge nodes and one flavour group attached to the ninth node) and $z^*=1$ (the non-dynamical external quark-antiquark pair transform in the fundamental of the first gauge node). As we stretch the separation $L$ in the $x$ direction, the F1 string explores both the $r$ and $z$ directions, turning around at the positions $r_0$ and $z_0$ respectively. 
Figure \ref{fig:exp0rz} shows $L$ (the separation in $x$ when $z=z^*$ and $r=\infty$) and the energy $E$ as functions of $r_0$ and $z_0$. We note that some phenomenon is happening when $z_0\sim (P-1)=9$, which is close to the position of the flavour branes.
\begin{figure}[!htbp]
\centering
 \includegraphics[width=.47\textwidth]{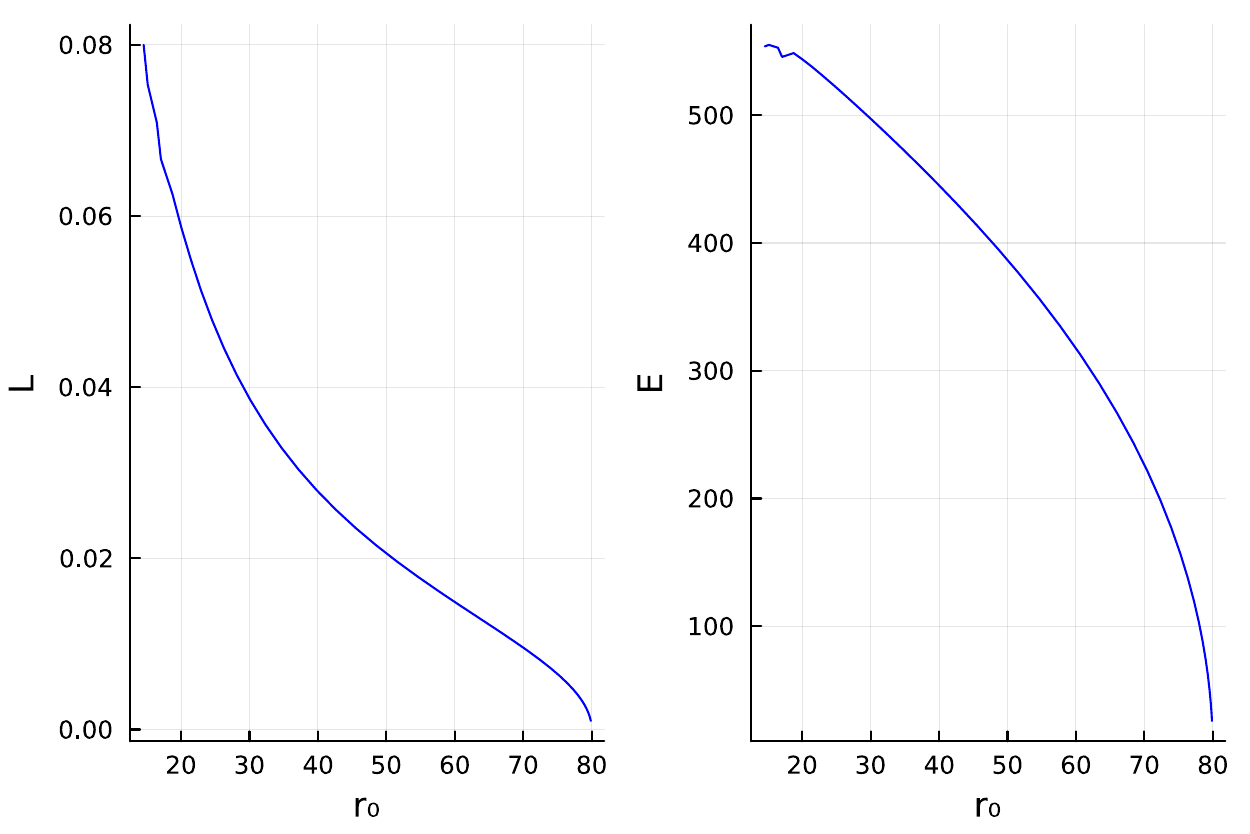}
\qquad
 \includegraphics[width=.47\textwidth]{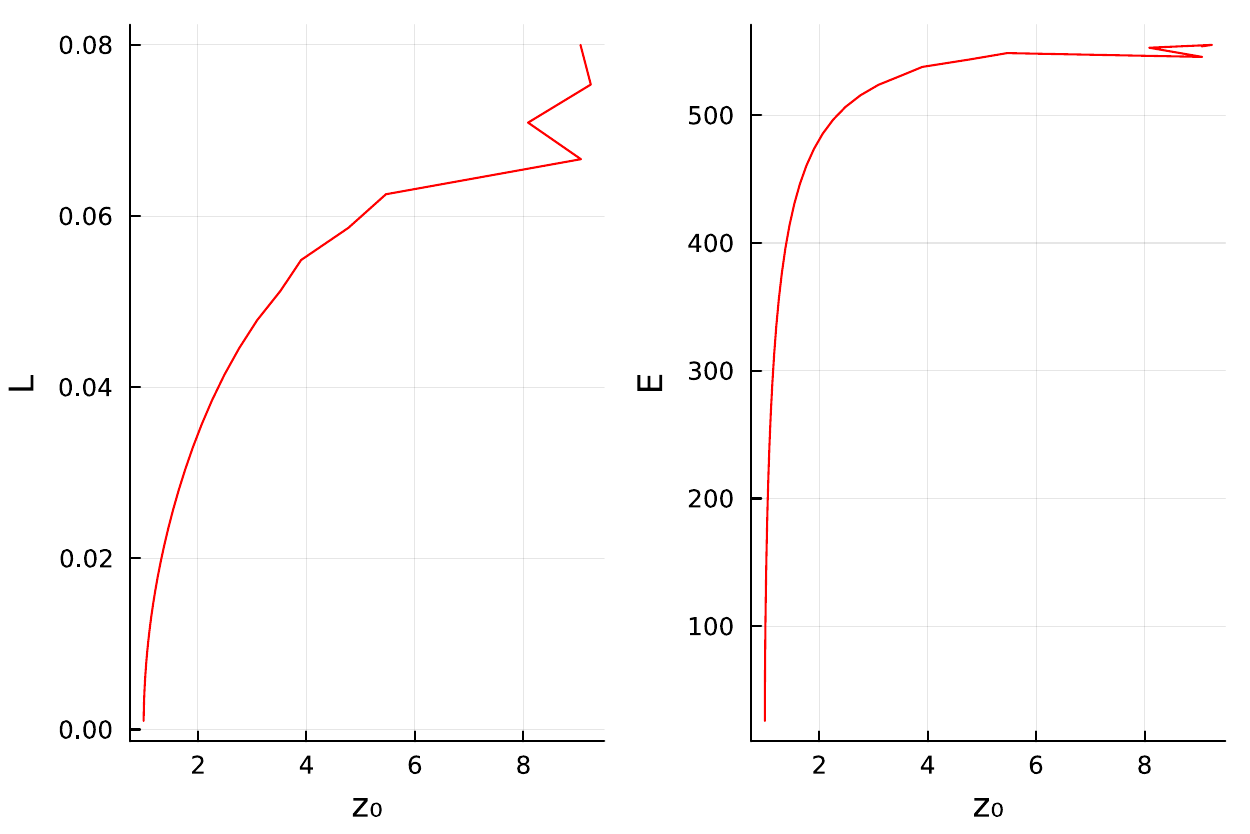}
\caption{The string separation $L$ and energy $E$ as functions of the ``tip'' of the sting in the $r$ and $z$ directions. Even with some numerical noise, a transition is clearly visible in the $L(z_0)$ plot.\label{fig:exp0rz}}
\end{figure}

Figure \ref{fig:exp0string} shows that, as we stretch the string in $x$, the profile explores regions closer to the source branes, staying close to $z\sim (P-1)=9$ for separations larger than a given critical distance $L_{crit}$.
\begin{figure}[!htbp]
\centering
 \includegraphics[width=.70\textwidth]{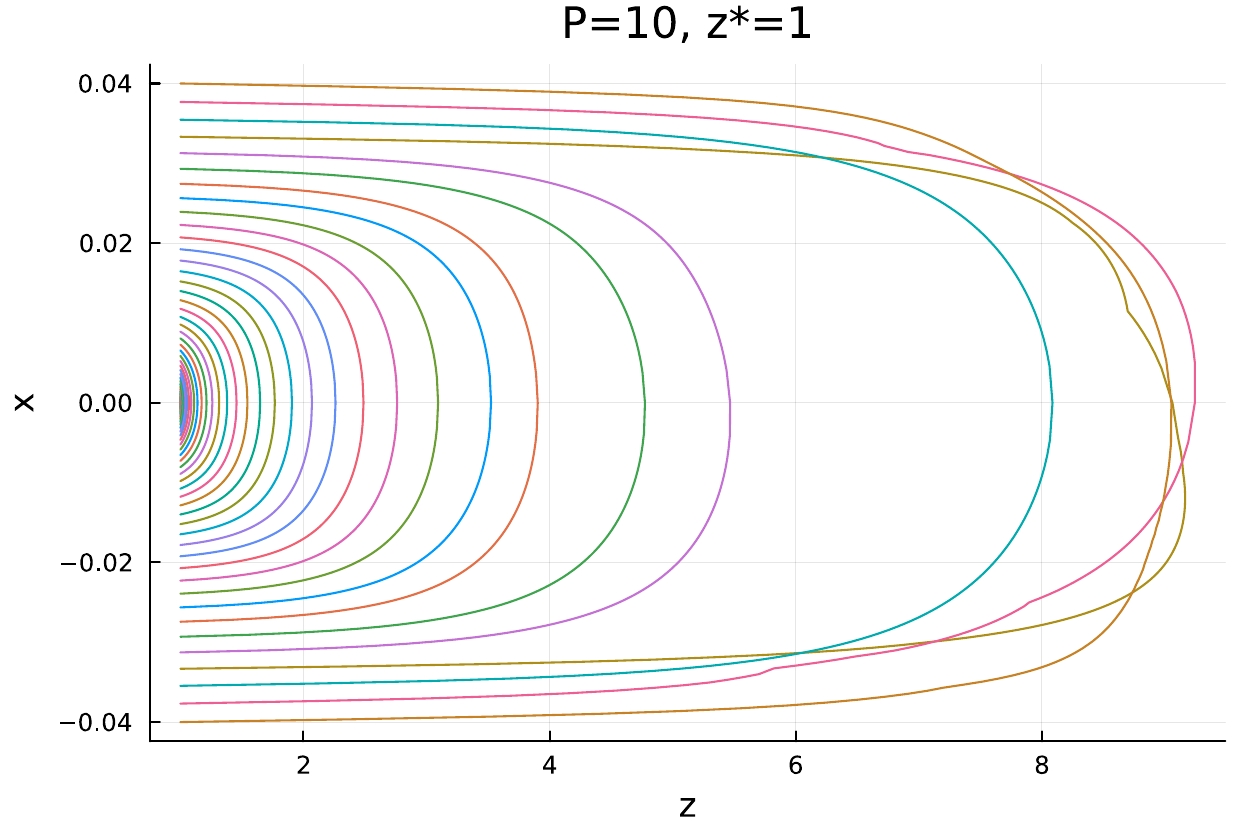}
\caption{The strings, extending in the $z$ direction, for various values of the separation $L$. It is visible that, after a certain separation, the string tends to stay close to the flavour group situated at $z=P-1$.\label{fig:exp0string}}
\end{figure}

Finally, Figure \ref{fig:exp0LE} is very interesting, as it displays that a linear (confining) behaviour for $E(L)$ is observed, which crosses-over to a screened behaviour as we increase $L$. Importantly, the larger $P$, the longer the range for which the linear (confining) behaviour $E\sim \sigma L$ is observed. This is in agreement with our intuition, that creating the operator in Figure \ref{fig:screeningquiver} becomes energetically more costly.
\begin{figure}[!htbp]
\centering
 \includegraphics[width=.70\textwidth]{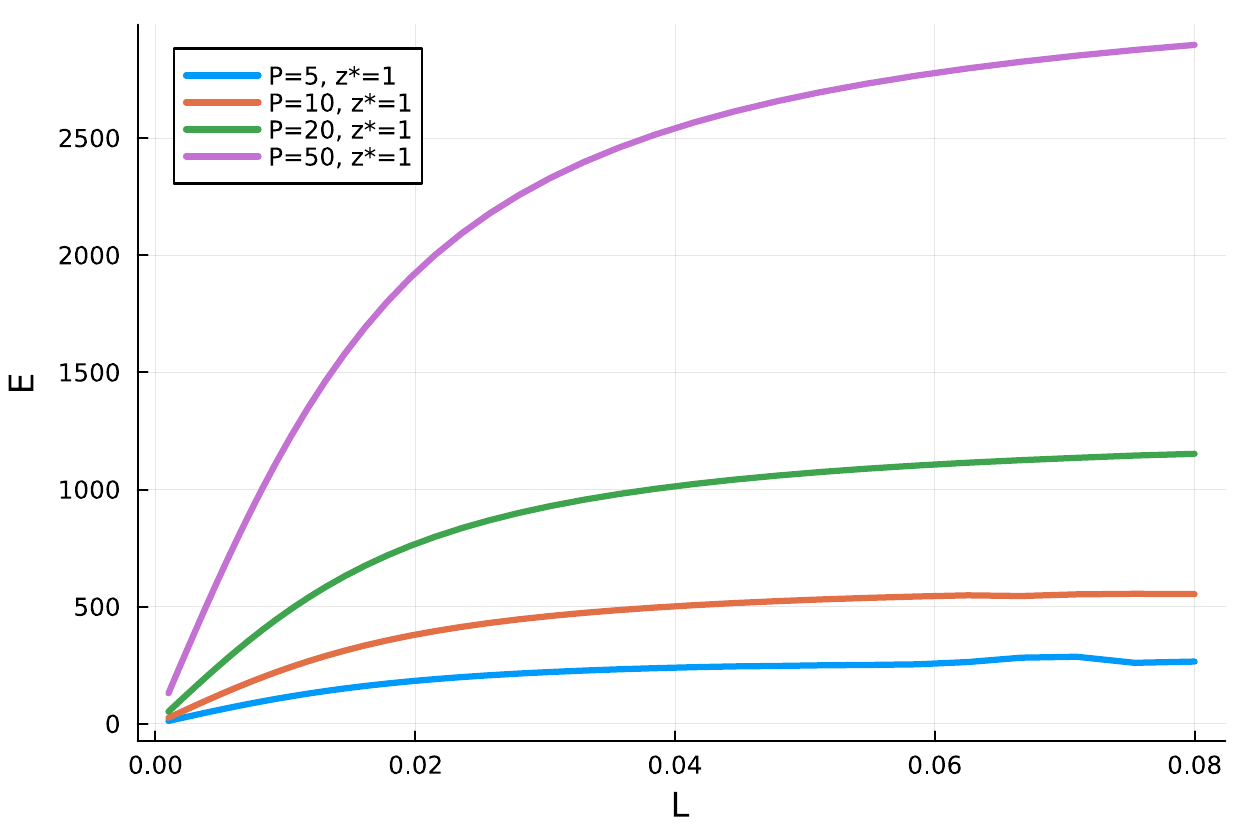}
\caption{The string energy $E$ as a function of the separation $L$, for various values of $P$, to analyze various distances to the flavour group. For higher values of $P$, the confining behavior holds for larger separations $L$.\label{fig:exp0LE}}
\end{figure}

We can find a fit function which estimates the energy of the quark-antiquark pair as the function of the separation $L$. Our proposal is a function of the form
\begin{equation} \label{eq:fit}
 E(L)=-\frac{a}{L}+\gamma\frac{1-e^{-b L}}{b}.
\end{equation}
For small $L$ ($L\ll b^{-1}$)
\begin{equation} \label{eq:fit2}
 E(L\rightarrow 0)=-\frac{a}{L}+\gamma L.
\end{equation}
which contains the conformal behaviour at small $L$ and transition towards confining with quark-antiquark string tension given by $\gamma$ parameter. For large $L$ we obtain
\begin{equation}
 E(L\rightarrow\infty)=\frac{\gamma}{b},
\end{equation}
which captures the screening behaviour as the energy function saturates. The potential function proposed here is inspired by the analysis performed in Ref.~\cite{Suganuma:1993ps}. In this work, the confining potential of QCD in the quenched quark approximation is derived using the abelian gauge fixing method and monopole condensation. By introducing dynamical light quarks one can introduce screening and derive an effective potential for the quark-antiquark pair which has a form similar to eq.\eqref{eq:fit}.

Approximating the data of Figure \ref{fig:exp0LE} for $P=10,20$ with the function in eq.\eqref{eq:fit}, we obtain the plot in Figure \ref{fig:fit1}. 
The derived values for the parameters are found as $a=0.011697, b= 54.766$ and $\gamma=30982.0$ for the quiver with $P=10$. Also, $a=0.0204072, b=53.0842$ and $\gamma=61542.3$ for the quiver with $P=20$. It is noteworthy that fitting a function of the conformal to confining transition form like $E(L)=-a/L+bL+c$ fails to give an appropriate approximation in all of the cases considered. This is a good indicator the we are actually observing a screening behaviour.

\begin{figure}[!htbp]
\centering
 \includegraphics[width=.80\textwidth]{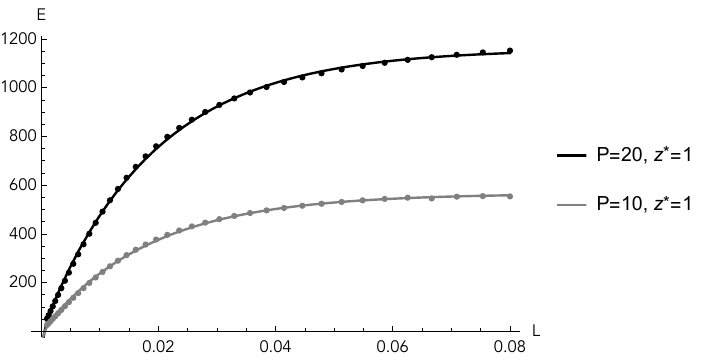}
\caption{Approximation of the data of Figure \ref{fig:exp0LE} for $P=10$ and $20$ with the function given in eq.\eqref{eq:fit}. 
The derived values for the parameters in the fitting functions are $a=0.011697, b= 54.766$ and $\gamma=30982.0$ for the quiver with $P=10$ and $a=0.0204072, b=53.0842$ and $\gamma=61542.3$ for the $P=20$ quiver.\label{fig:fit1}}
\end{figure}

\FloatBarrier
\subsubsection{Experiment 2: isosceles triangle rank}

For the second quiver, we chose one that features a flavour group at $z=P/2$, resulting in a isosceles triangular rank function, as it is visible in Figure \ref{fig:exp1alpha}.

\begin{figure}[!htbp]
\centering
 \includegraphics[width=.50\textwidth]{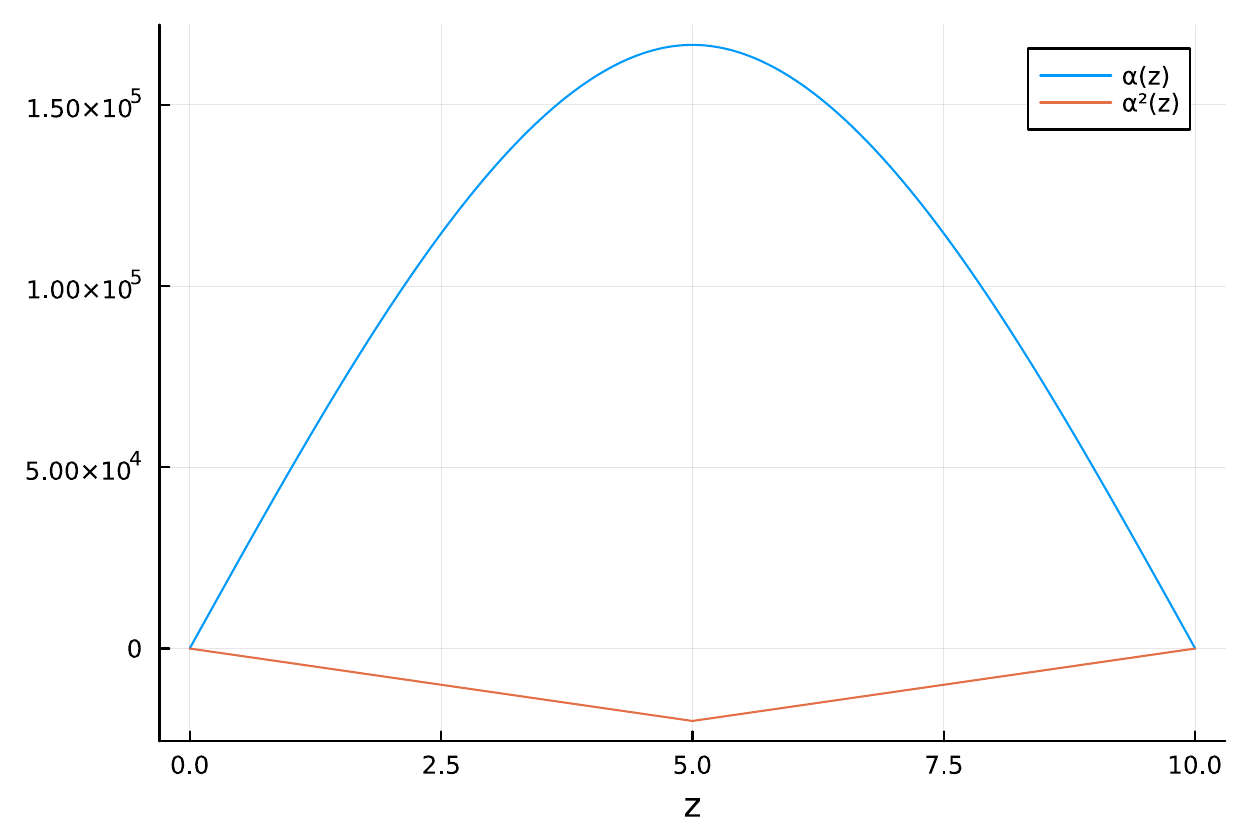}
\caption{The $\alpha(z)$ and $\alpha''(z)$ functions for the isosceles triangular quiver. The rank function presents a kink at $z=P/2$, the position of the flavour group.\label{fig:exp1alpha}}
\end{figure}

In fact, for 
$$
R(z)=-\frac{1}{81\pi^2}\alpha''(z) = \left\{
 \begin{array}{ll}
N z & \quad 0 \leq z\leq (P/2), \\
N (P-z) & \quad (P/2)\leq z\leq P;
 \end{array}
 \right.
$$
the corresponding quiver is
\begin{center}
	\begin{tikzpicture}
	\node (1) at (-6,0) [circle,draw,thick,minimum size=1.2cm] {N};
	\node (2) at (-4,0) [circle,draw,thick,minimum size=1.2cm] {2N};	
	\node (3) at (-2,0) {$\dots$};
 \node (4) at (0,0) [circle,draw,thick,minimum size=1.2cm] {PN/2};
 \node (5) at (2,0) {$\dots$};
	\node (6) at (4,0) {(P-1)N};
 \node (7) at (6,0) [circle,draw,thick,minimum size=1.2cm] {PN};
 \node (4q) at (0,-2) [rectangle,draw,thick,minimum size=1.2cm] {2N};
	\draw[thick] (1) -- (2) -- (3) -- (4) -- (5)-- (6) -- (7);
	\draw[thick] (4,0) circle (0.7cm) ;
	\draw[thick] (1,0) -- (1.3,0);
	\draw[thick] (2.7,0) -- (3.3,0);
 \draw[thick] (4) -- (4q);
	\end{tikzpicture}\
\end{center}

We set $P=10$, and vary the position of the gauge node for which we compute the Wilson loop, changing $z^*$. We expect the string to extend from the chosen $z^*$ towards the kink at $z=P/2$, and the confining behavior to have a shorter range the closest we set $z^*$ to $P/2$.
The results of this analysis are visible in Figures \ref{fig:exp1rz}, \ref{fig:exp1string} and \ref{fig:exp1LE}.
\\
In the same fashion as in the previous experiment, as we stretch the separation $L$ in the $x$ direction, the string explores both the $r$ and $z$ directions, turning around at the positions $r_0$ and $z_0$ respectively. 
Figure \ref{fig:exp1rz} shows $L$ (the separation in $x$ when $z=z^*$ and $r=\infty$) and the energy $E$ as functions of $r_0$ and $z_0$. We notice that, as we increase $L$, the value of $z_0$ does get closer to $P/2$, but eventually overshoots it. Comments regarding this behavior are presented in Section \ref{sec:discussion}: the behavior for strings that extend beyond the closest flavour group is not quantitatively well-described by our approach, but we can still analyze their qualitative working.
\begin{figure}[!htbp]
\centering
 \includegraphics[width=.47\textwidth]{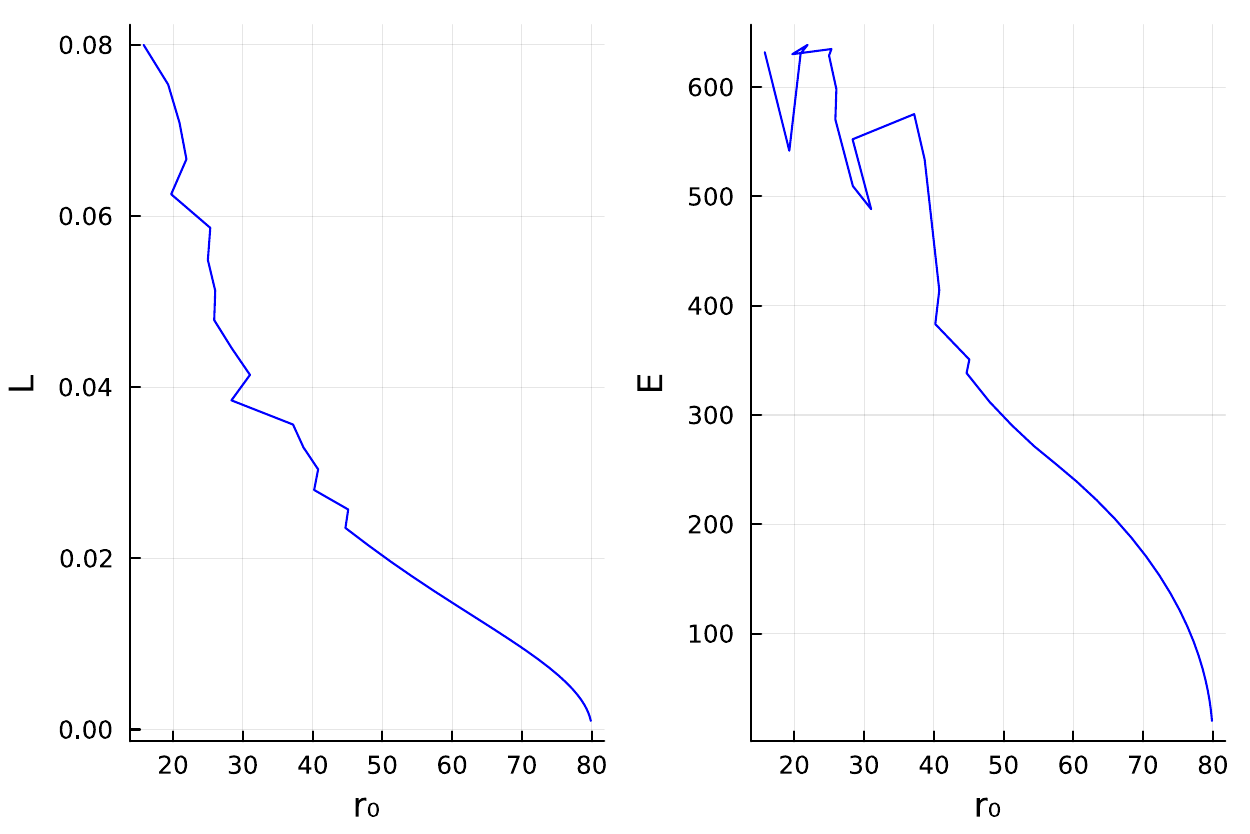}
\qquad
 \includegraphics[width=.47\textwidth]{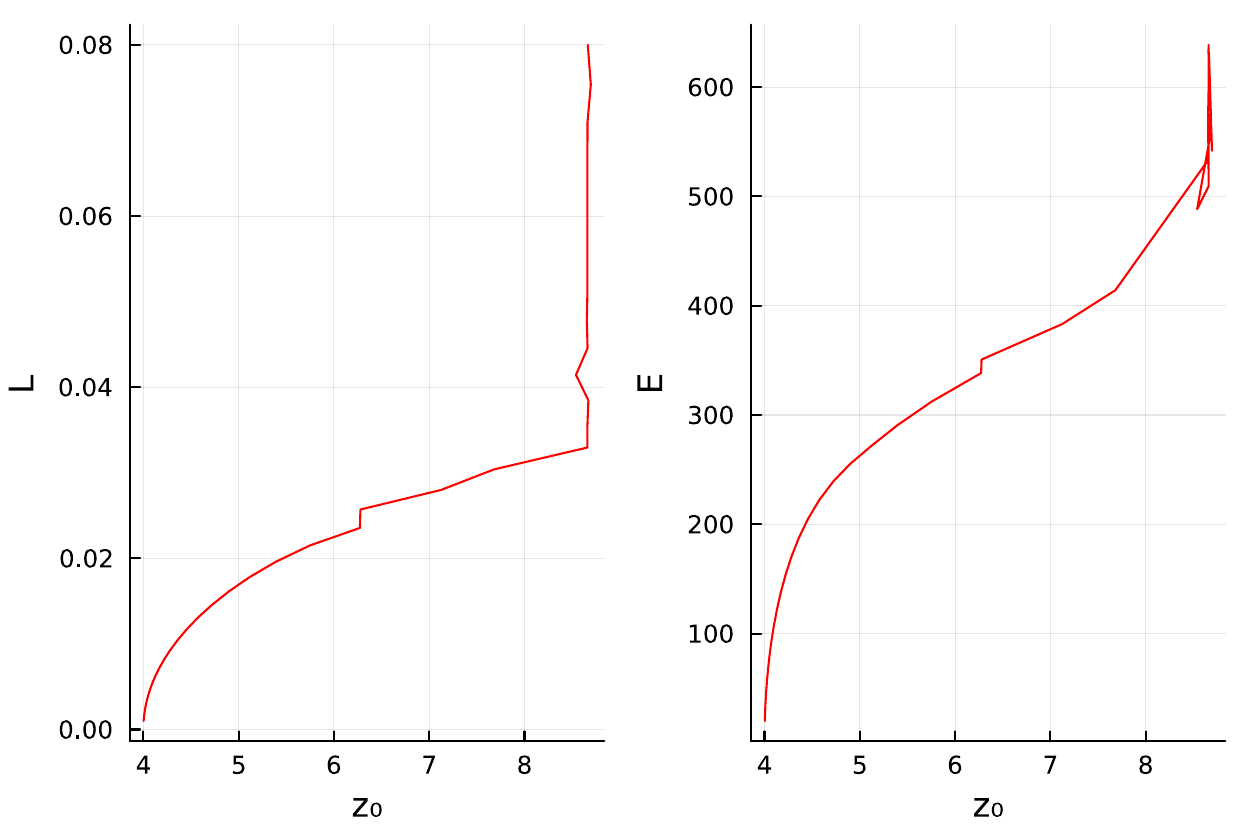}
\caption{The string separation $L$ and energy $E$ as functions of the ``tip'' of the sting in the $r$ and $z$ directions. In the above plots, $z^*=4$ and the closest flavour node $z_{cf}=5$. The transition is still clearly visible in the $L(z_0)$ plot.\label{fig:exp1rz}}
\end{figure}

Figure \ref{fig:exp1string} shows that, as we stretch the string in $x$, the profile explores regions closer to the source branes, staying close to $z\sim (P-1)=9$ for separations larger than a given critical distance $L_{crit}$. Figure \ref{fig:exp1LE} still shows the linear (confining) behaviour for $E(L)$ that crosses into a screened behaviour. 
\begin{figure}[!htbp]
\centering
 \includegraphics[width=.70\textwidth]{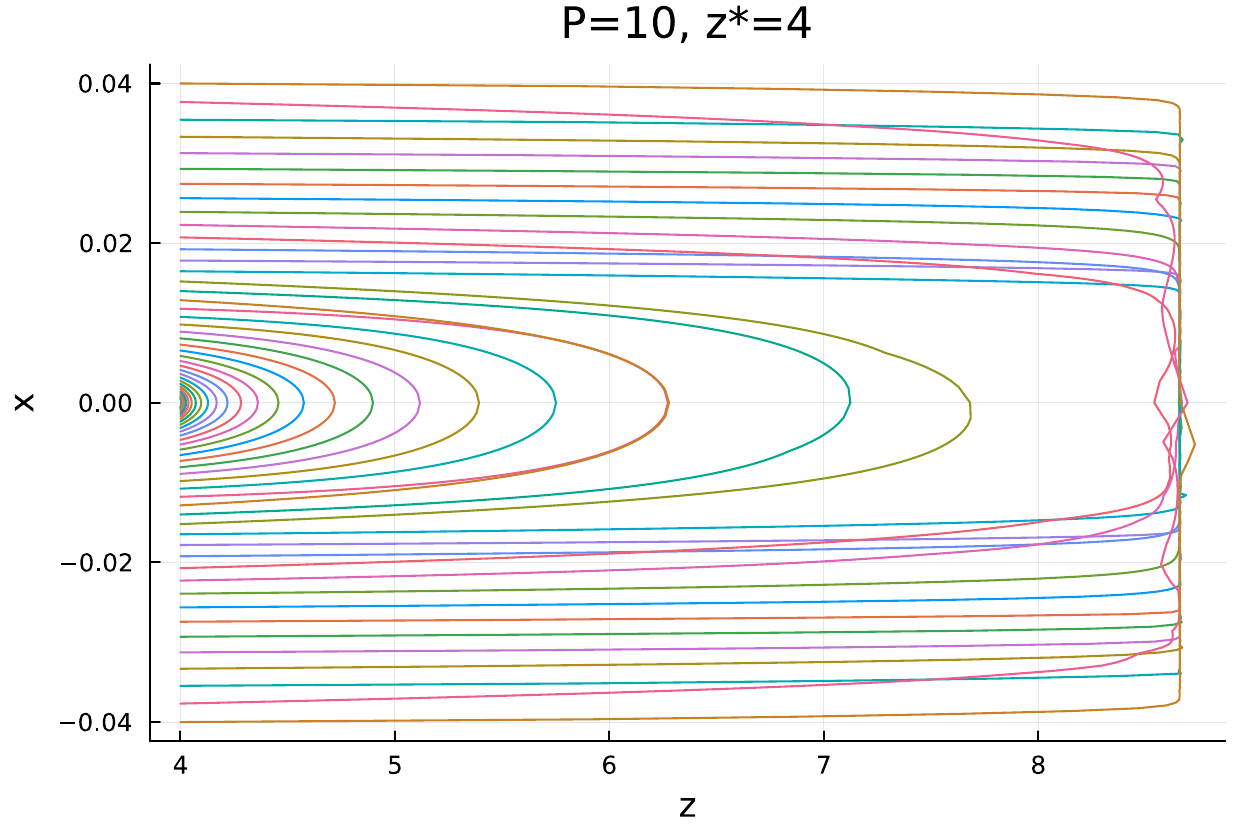}
\caption{The strings, extending in the $z$ direction, for various values of the separation $L$. It is even more visible that, after a certain separation, the string does not feature a ``U-shape'' and instead tends to dive into the $z$ direction and ``stick'' to a particular value of $z$.\label{fig:exp1string}}
\end{figure}
\begin{figure}[!htbp]
\centering
 \includegraphics[width=.70\textwidth]{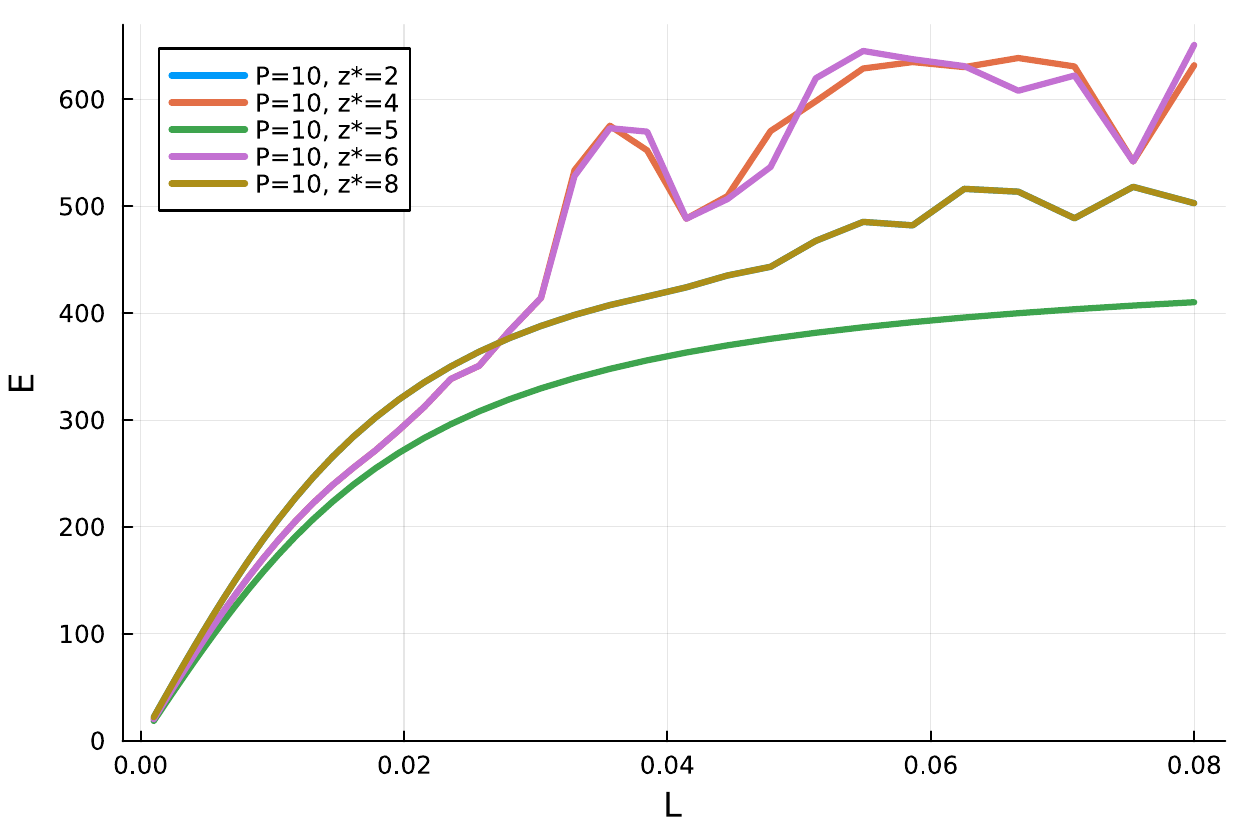}
\caption{The string energy $E$ as a function of the separation $L$, for various values of $z^*$, to analyze various distances to the flavour group. For $L>L_{crit}$, the critical value of the separation after which the ``U-shaped'' strings are not preferred anymore, the results are very noisy, except for the string already situated at the flavour group, with $z^*=P/2=5$: for that case, the $z$-dynamics does not depend on $L$, and the energy $E(L)$ only depends on the extension in the $r$ direction.\label{fig:exp1LE}}
\end{figure}

The resulting plot for approximating the data of Figure \ref{fig:exp1LE} for $P=10,\, z^*=5$ with the function in eq.\eqref{eq:fit} is given in Figure \ref{fig:fit2}. 
The derived values for the parameters are found as $a=0.00709182, b= 52.8363$ and $\gamma=21771.7$. Again, the screening behaviour is confirmed in this case.
\begin{figure}[!htbp]
\centering
 \includegraphics[width=.80\textwidth]{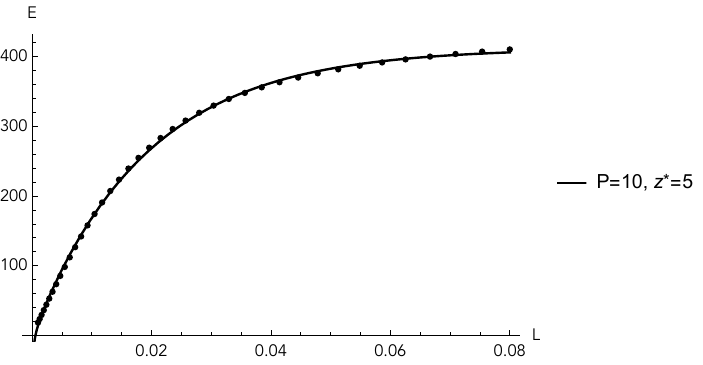}
\caption{Approximation of the data of Figure \ref{fig:exp1LE} for $P=10,\, z^*=5$ with the function given in eq.\eqref{eq:fit}. 
The derived values for the parameters in the fitting functions are $a=0.00709182, b= 52.8363$ and $\gamma=21771.7$.\label{fig:fit2}}
\end{figure}

\FloatBarrier
\subsubsection{Experiment 3: isosceles trapezoid rank}

Here, we choose a quiver that features two flavour groups, one at $z=1$ and one at $z=P-1$, resulting in an isosceles trapezoidal rank function, as it is visible in Figure \ref{fig:exp3alpha}. 
Indeed, 
for 
$$
R(z)=-\frac{1}{81\pi^2}\alpha''(z) = \left\{
 \begin{array}{ll}
N z & \quad 0 \leq z\leq 1, \\
N & \quad 1\leq z\leq (P-1);\\
N (P-z) & \quad (P-1)\leq z\leq P;
 \end{array}
 \right.
$$
the corresponding quiver is
\begin{center}
	\begin{tikzpicture}
 \node (0) at (-8,0) [rectangle,draw,thick,minimum size=1.2cm] {N};
	\node (1) at (-6,0) [circle,draw,thick,minimum size=1.2cm] {N};
	\node (2) at (-4,0) [circle,draw,thick,minimum size=1.2cm] {N};
	\node (3) at (-2,0) [circle,draw,thick,minimum size=1.2cm] {N};	
	\node (4) at (0,0) {$\dots$};
	\node (6) at (4,0) [rectangle,draw,thick,minimum size=1.2cm] {N};
	\node (5) at (2,0) [circle,draw,thick,minimum size=1.2cm] {N};
	\draw[thick] (0) -- (1) -- (2) -- (3) -- (4) -- (5)-- (6);
	\draw[thick] (1,0) -- (1.3,0);
	\draw[thick] (2.7,0) -- (3.3,0);
	\end{tikzpicture}\
\end{center}
We study this quiver to learn how a string behaves when there are two possible flavour groups to reach. Our physical intuition tells us that each quark-antiquark probe will tend to the closest flavour node. The behaviour for equidistant flavour groups (i.e. $z^*=P/2$) is nontrivial and is analyzed in Figure \ref{fig:exp3string}.

\begin{figure}[!htbp]
\centering
 \includegraphics[width=.60\textwidth]{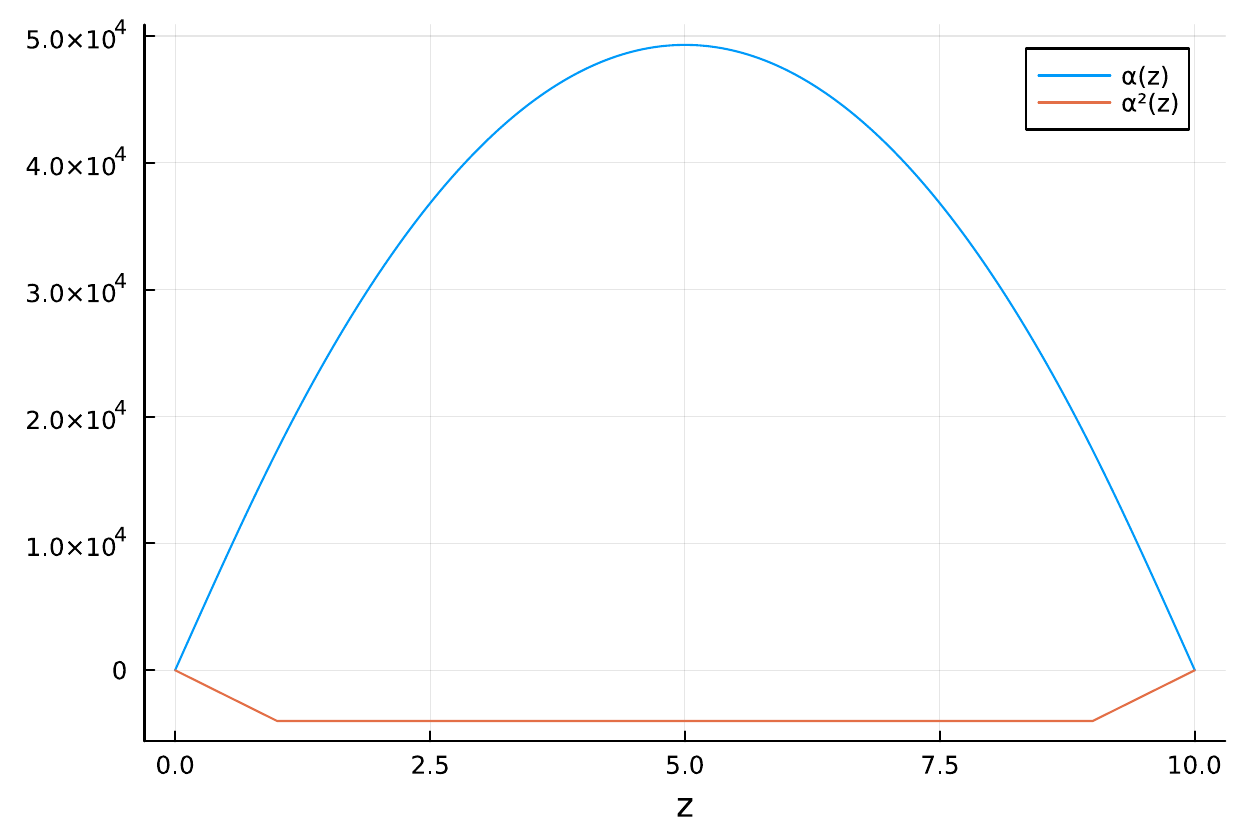}
\caption{The $\alpha(z)$ and $\alpha''(z)$ functions for the isosceles trapezoidal quiver. The rank functions has two kinks, at $z=1$ and $z=P-1$.\label{fig:exp3alpha}}
\end{figure}

We set our Wilson loop in the various nodes, including the middle one, setting $z^*$ accordingly.
The results of this analysis can be seen in Figures \ref{fig:exp3rz}, \ref{fig:exp3string} and \ref{fig:exp3LE}.
\\
We set $P=10$.
Similar to the previous cases, as we stretch the separation $L$ in the $x$ direction, the F1 string explores both the $r$ and $z$ directions, turning around at the positions $r_0$ and $z_0$, respectively. 
Figure \ref{fig:exp3rz} shows $L$ (the separation in $x$ when $z=z^*$ and $r=\infty$) and the energy $E$ as functions of $r_0$ and $z_0$. %
\begin{figure}[!htbp]
\centering
 \includegraphics[width=.47\textwidth]{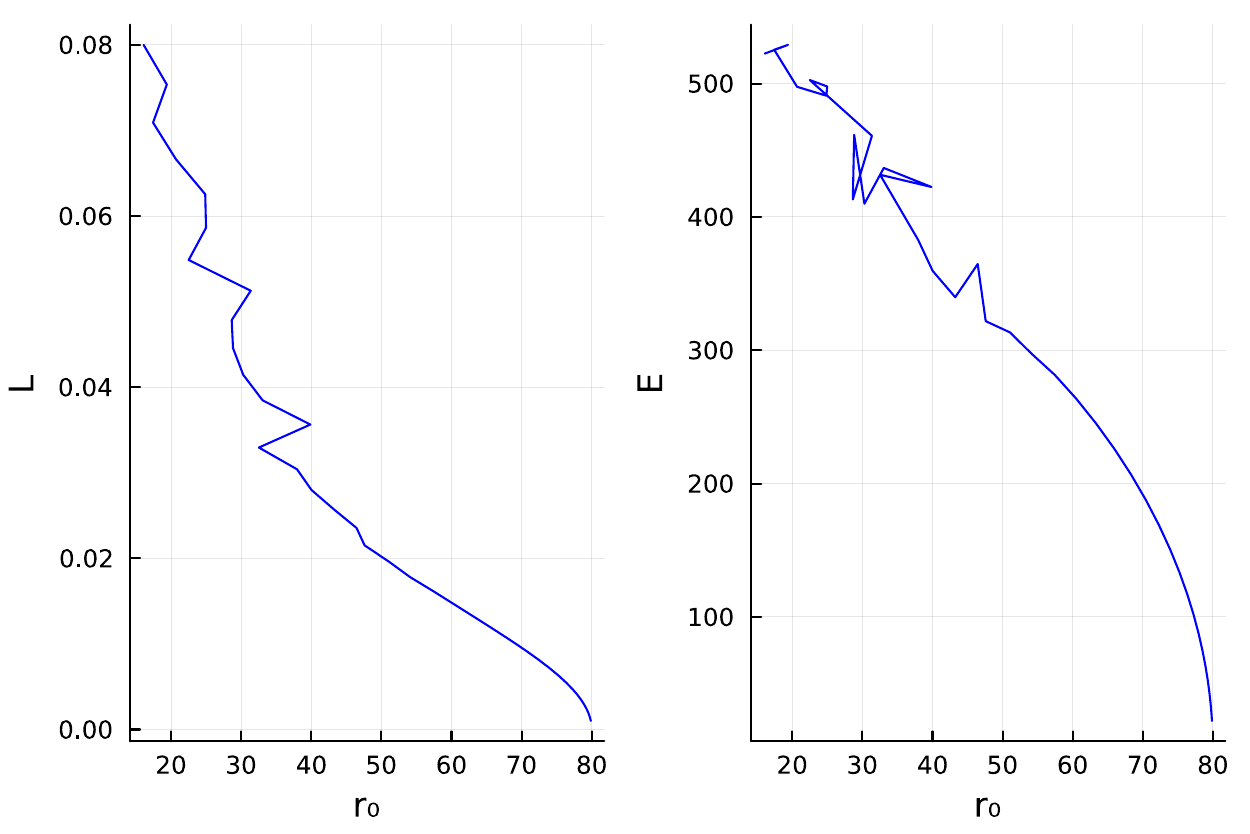}
\qquad
 \includegraphics[width=.47\textwidth]{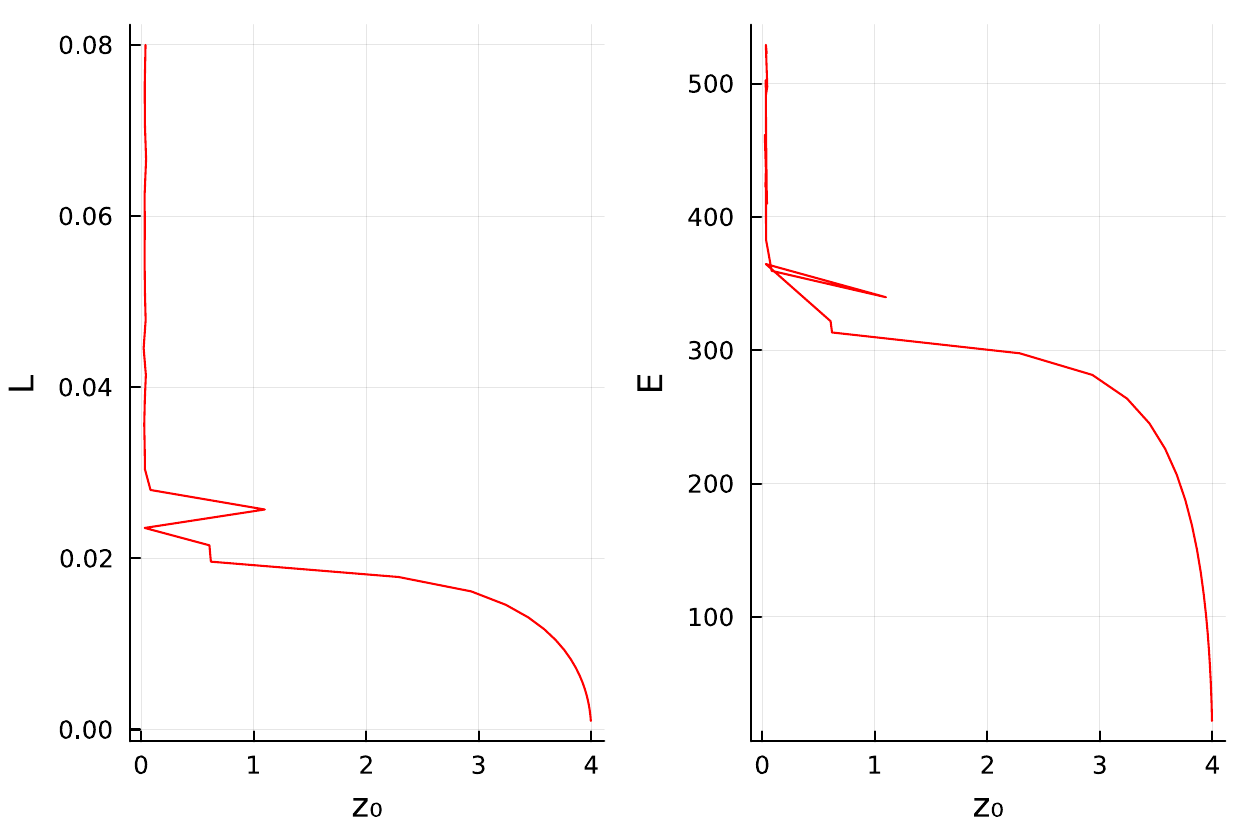}
\caption{The string separation $L$ and energy $E$ as functions of the ``tip'' of the sting in the $r$ and $z$ directions. Compared to the previous cases, the appearance of the $L(z_0)$ and $E(z_0)$ functions is mirrored, as the closest flavour group to $z^*=4$ is at a lower value of $z$.\label{fig:exp3rz}}
\end{figure}

Inspecting Figure \ref{fig:exp3string}, we observe that when $z^*=P/2=5$, there are two equivalent choices of flavour branes for the string to attach to. Hence the string remains in the middle of the quiver at $z^*=P/2=5$ and can not decide between the two. Fluctuations of the string will eventually lead to a choice for the string to fall on one of the branes either on the left or the right flavour groups.

\begin{figure}[!htbp]
\centering
 \includegraphics[width=.80\textwidth]{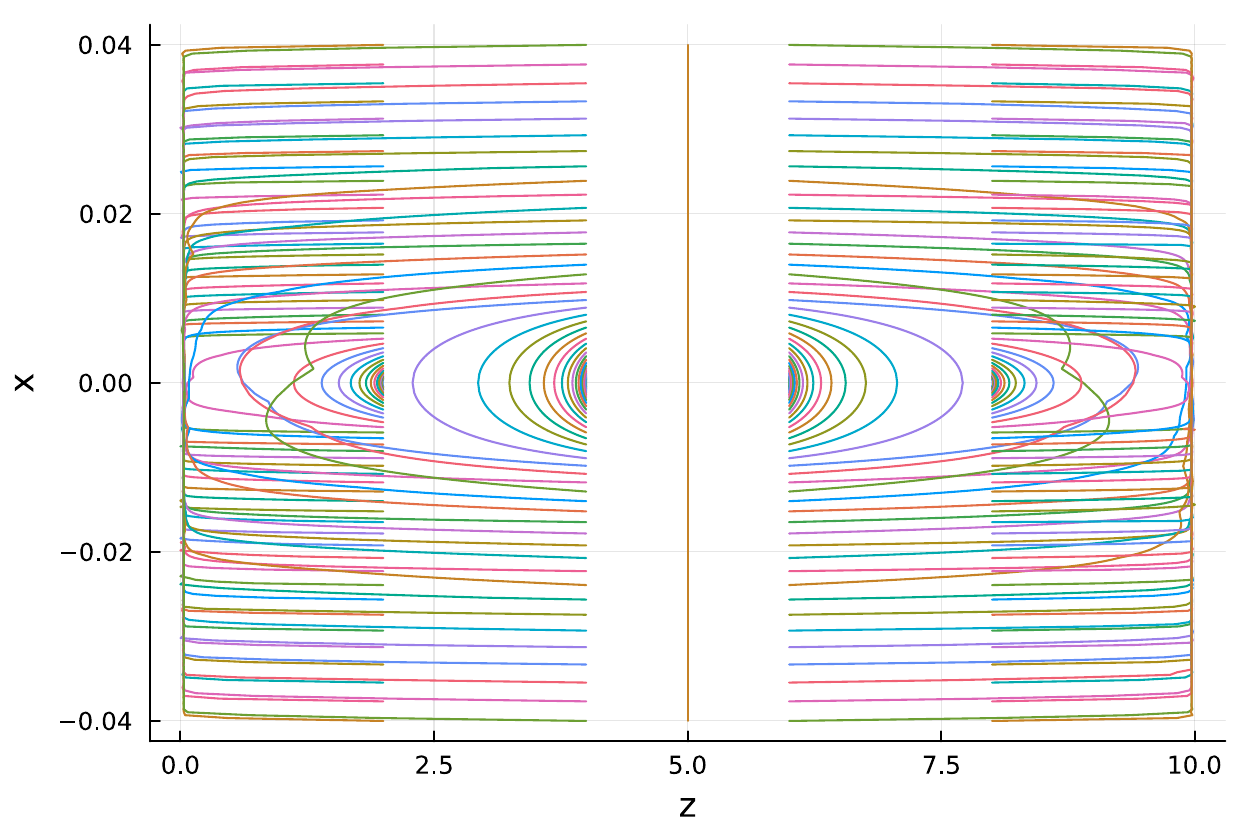}
\caption{The strings, extending in the $z$ direction, for various values of the separation $L$ and of the position of the probe $z^*$. Clearly, strings with $z^*<P/2$ dive down to $z=0$, while if $z^*>P/2$ the strings tends to $z=10$. The nontrivial case of $z^*=P/2$ shows a string that, with no perturbations, holds its position.\label{fig:exp3string}}
\end{figure}
Figure \ref{fig:exp3LE} shows similar behaviour to the previous cases.
\begin{figure}[!htbp]
\centering
 \includegraphics[width=.80\textwidth]{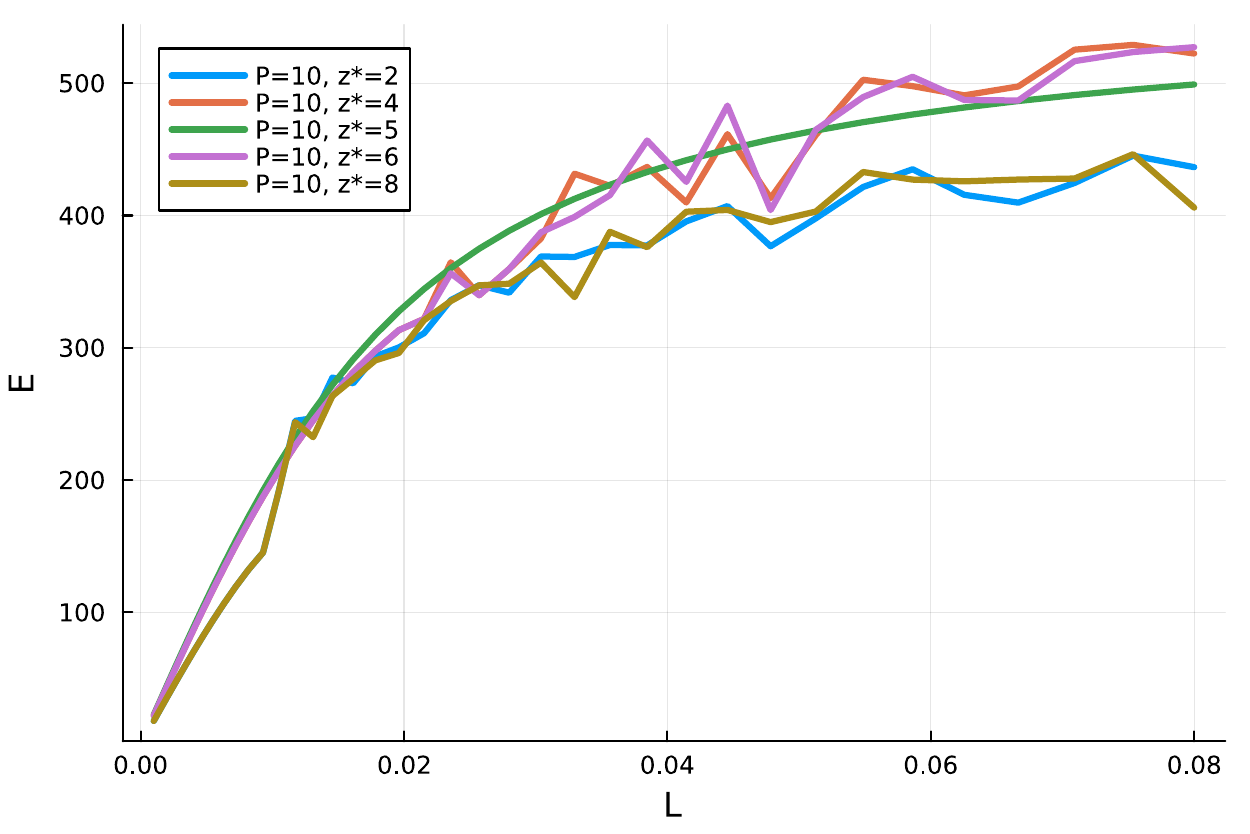}
\caption{The string energy $E$ as a function of the separation $L$, for various values of $z^*$.\label{fig:exp3LE}}
\end{figure}
The similar resulting plot for approximating the data of Figure \ref{fig:exp3LE} for $P=10, \,z^*=5$ with the function in eq.\eqref{eq:fit} is given in Figure \ref{fig:fit3}.
The derived values for the parameters are found as $a=0.00862758, b=52.8362$ and $\gamma=26486.4$.

\begin{figure}[!htbp]
\centering
 \includegraphics[width=.80\textwidth]{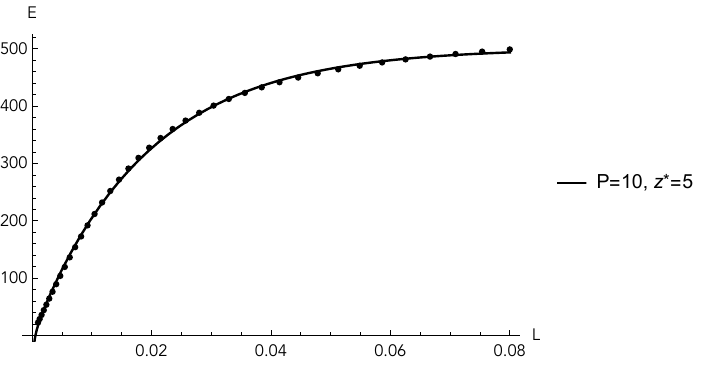}
\caption{Approximation of the data of Figure \ref{fig:exp3LE} for $P=10, \, z^*=5$ with the function given in eq.\eqref{eq:fit}. 
The derived values for the parameters in the fitting functions are $a=0.00862758, b=52.8362$ and $\gamma=26486.4$.\label{fig:fit3}}
\end{figure}
\subsection{Discussion of the results}\label{sec:discussion}

In this section, we comment on various aspects of the results presented above. 

Let us start with a conceptual observation. As we are working with linear quivers, some of our gauge nodes are not directly connected to a flavour group. In this sense, the systems we studied are different from a canonical example like QCD (one gauge node connected to one flavour group). Our `screening process' is suppressed with respect to the usual screening, as it involves exciting an operator like the one we write in eq.\eqref{operatorred}. In a QCD-like theory, screening is associated with the presence of particles transforming non-trivially under the center of the gauge group. 
Also, for the case of a generic linear quiver with flavours, the center symmetry is likely broken\footnote{Thanks to Jeremias Aguilera-Damia for an explanation on this.}.

Regarding the creation of the dynamical quark-antiquark pair, we already mentioned that the process is not kinematically suppressed in our set-ups. In fact, the snapping of the U-shaped connected string into two strings (connected to the dynamical quark-antiquark pair) is proportional to $g_s\sim \frac{1}{N_c}$ (being $N_c$ the rank of the gauge node on which we insert the non-dynamical quark pair). This is enhanced by the number of possibilities for the string to connect to a flavour brane, hence the process is weighted by $\frac{N_f}{N_c}\sim 1$ (here $N_f$ is the rank of the global symmetry group). 

It is important to emphasise that our Nambu-Goto action does not contemplate the actual screening process. We are not finding a solution where the fundamental U-shaped string actually disconnects and reconnects to the flavour branes (sources) present in the background. What we really observe is that, as the U-shaped string explores regions close to the sources, a qualitative change in the shape of the string occurs, and a change in the slope of the $E(L)$ curve takes place. We interpret these changes as screening. 

As this change happens, and the string tries to explore the region even beyond the sources, the numerical algorithm fails to converge to a smooth solution. The string keeps extending as it was ``pulled'' beyond the source, towards the boundaries of the $z$ direction, $0$ and $P$, yielding the noisy results visible in all the plots of Section \ref{sec:solutions} for high values of $L$, especially in Figures \ref{fig:exp3LE}, \ref{fig:exp3rz} for which the critical value $L_{crit}$ is lower, and therefore there are more values of $L$ for which the strings surpass the source. In fact, we notice that for $L<L_{crit}$ the numerical error is fully under control, thus we know how the string extends towards the source. When instead $L>L_{crit}$, the obtained results cannot be trusted quantitatively, and we present them only for completeness, with no confidence of knowing how the string actually attaches to the source.

The screening process involves the competition between two configurations. In one configuration the string is U-shaped (connected), whilst in the second
the string disconnects in the $z$-direction and attaches to the flavour branes (sources). The screening occurs when the disconnected string is energetically favoured over the U-shaped one. Aside from this the `snapping' should not be kinematically suppressed. In our case, with $\frac{N_f}{N_c}\sim 1$, the process is of order one. Notice also that the string that extends along $z$ from the gauge group to the nearest flavour group (with fixed $r=\bar{r}$) is finite and with finite energy.

To have a numerical grasp on how the screening process happens, we can force the string to the $z$-value of the kink as soon as $z_0$ reaches it, as shown in Figure \ref{fig:snapped}. We regard this as an ``informed trick'': we are aware that it is not a proper solution, yet we know from numerical evidence and physical intuition that, when the string reaches the source, the phenomenon cannot be well-captured by the approach presented before. Still, even in this analysis, there is uncertainty as to how we shall correctly interpolate between the U-shaped and straight configurations: we leave this to future study.

\begin{figure}[!htbp]
\centering
 \includegraphics[width=.45\textwidth]{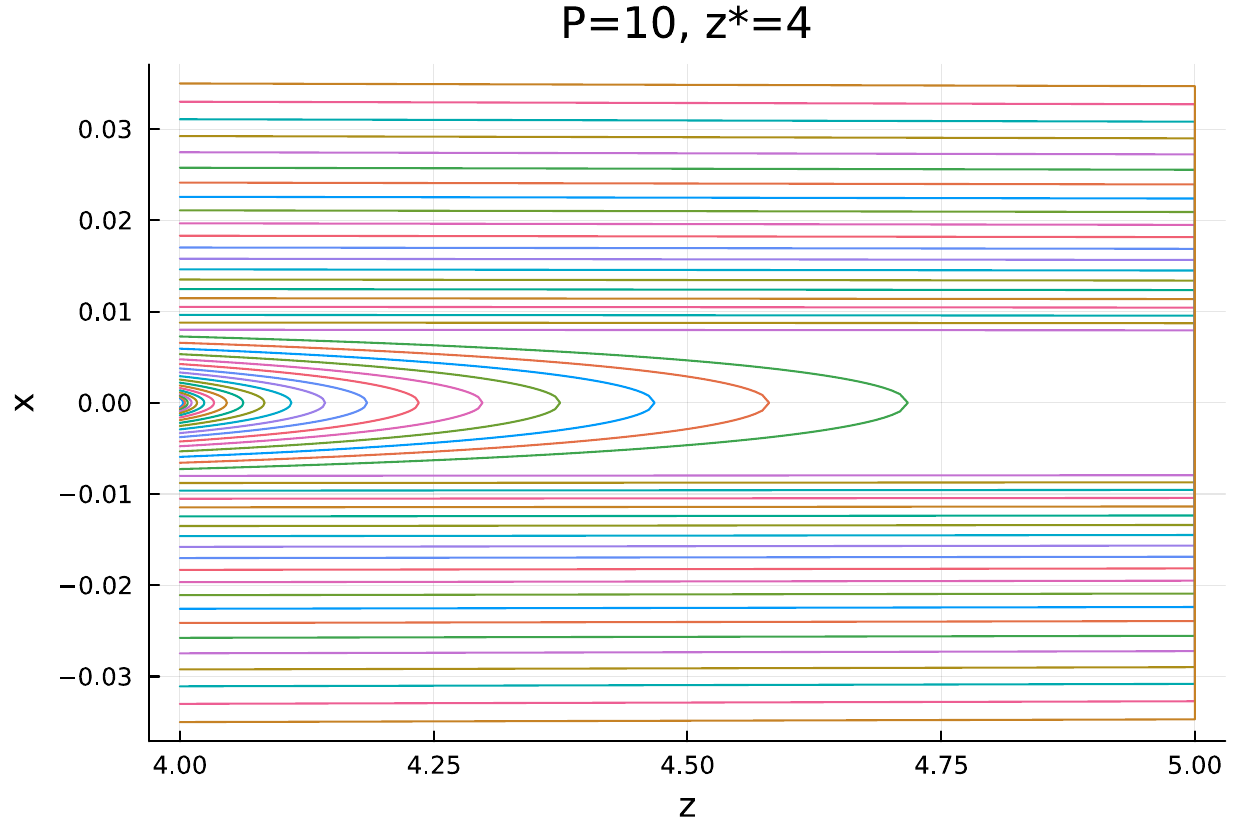}
 \includegraphics[width=.45\textwidth]{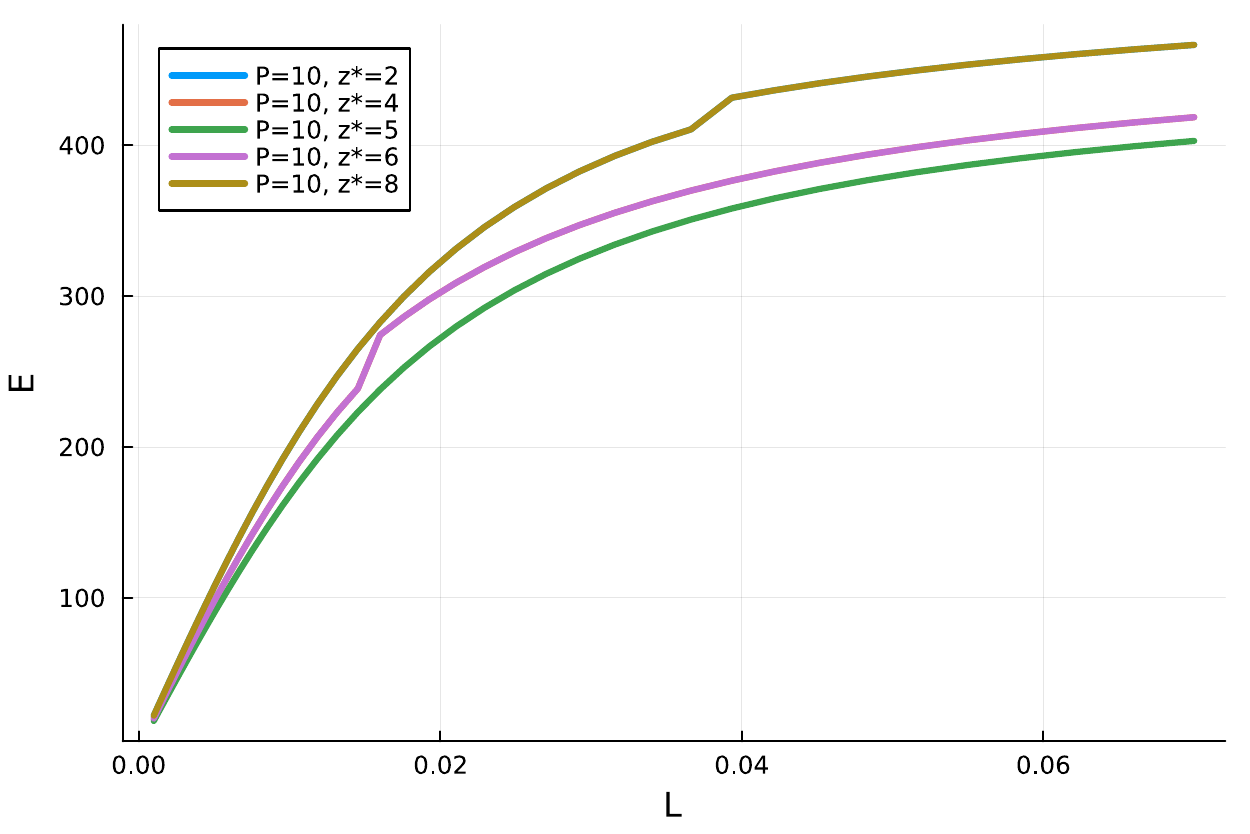}
\caption{The ``snapped'' version of Figures \ref{fig:exp1string} and \ref{fig:exp1LE}. The optimization in $r$ is left untouched, while the one in $z$ is modified: for all $L$ that would yield $z_0>z_{cf}$, the string ``snaps'', i.e. is fixed at $z(x)=z_{cf}\,\,\forall x$.\label{fig:snapped}}
\end{figure}

Another conceptual comment that should be made for full disclosure is the following: the low energy regime of the `mother' six dimensional SCFTs dual to the solution in eq.(\ref{backgroundads7xm3}) is Lagrangian. As this theory undergoes compactification, whose dual is the background in eq.(\ref{metric-dil-BPT}), the system becomes non-Lagrangian. When we write that the gauge nodes are located at integer positions of $z^*$ and the flavour nodes at the kinks of the rank function, we are abusing of language. The gauge/flavour nodes picture needs not be fully accurate in the 4D QFT.

It is interesting to comment briefly about the semi-analytic interpolation given by $E(L)$ in eq.(\ref{eq:fit}). We obtain this expression purely fitting our numerical results with a trial (analytic) function. Similar expressions were obtained in more phenomenological approaches, see for example \cite{Suganuma:1993ps}. It is of interest to analyse if the similarity between our expression (\ref{eq:fit}) and the expressions in \cite{Suganuma:1993ps} is beyond a coincidence. On the other hand, the fit we find is very accurate: using the Mean Absolute Percentage Error (MAPE) measure, we find that our fits following \eqref{eq:fit} in Figures \ref{fig:fit1}, \ref{fig:fit2}, \ref{fig:fit3} all have a MAPE under $3\%$, while for comparison, a fit of the form $E(L)=a/L+bL+c$ on the same data yields MAPEs higher than $30\%$.

Whilst we did not use this extensively in the above treatment, it is interesting to see the way in which the probe F1 explores the holographic space ($r,z$). This is what Figure \ref{strings3d} shows, for the first numerical experiment--see Section \ref{rankscalene}.
We can appreciate how the string starts exploring the radial direction, and at the same time starts to tilt slightly in the $z$ direction. There is probably interesting phenomenology hidden in the shape of these curves. We shall study this in the future.
\begin{figure}[!htbp]
\centering
 \includegraphics[width=.80\textwidth]{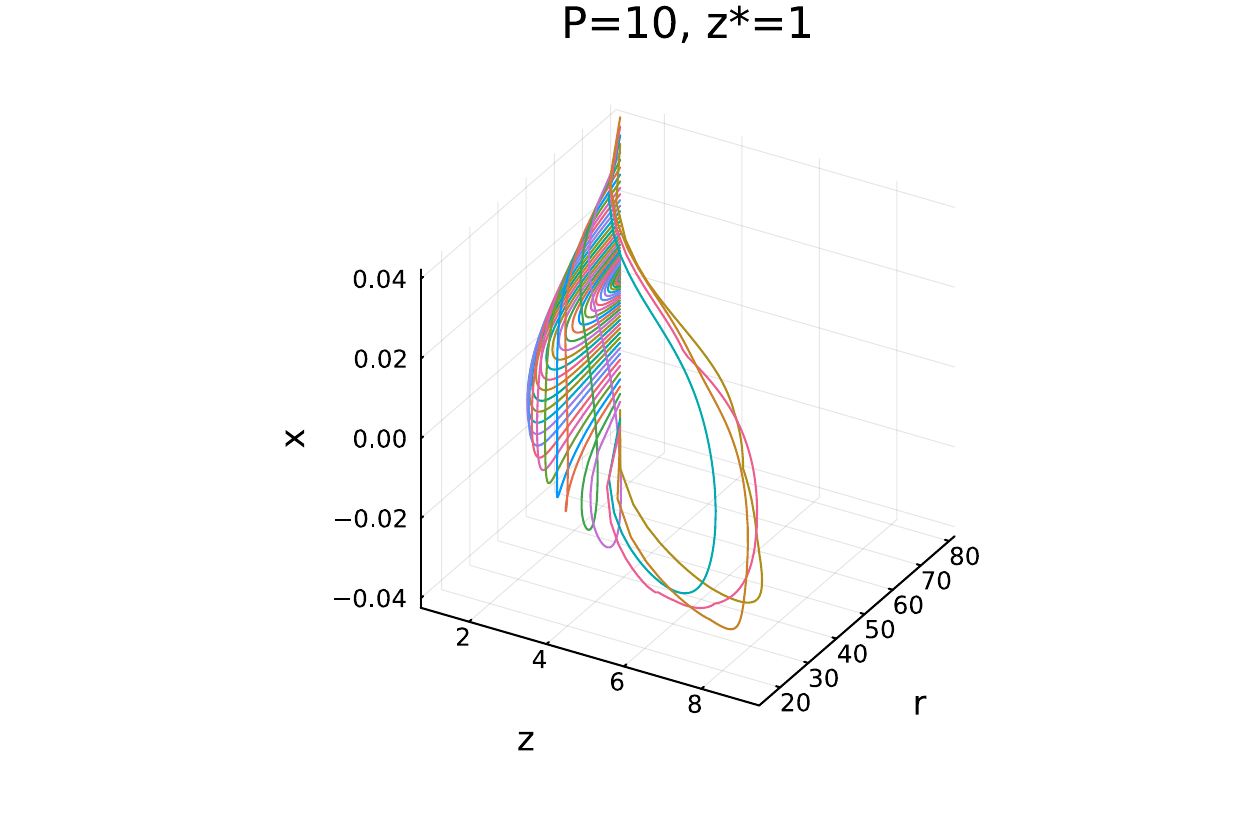}
\caption{The strings of the first experiment, already shown in Figure \ref{fig:exp0string}, extending in both the $z$ and $r$ directions, for various values of the separation $L$. It is visible how the strings bend in the $(r,z)$ plane.\label{fig:exp03D} }
\label{strings3d}
\end{figure}

\section{Conclusions and Closing Comments}\label{concl}
Let us present brief and general conclusions, and propose some topics that this papers suggest for future study. We start itemising some of the salient points of this paper.
\begin{itemize}
 \item{We described the construction of an infinite family of massive IIA backgrounds dual of a family of four dimensional ${\cal N}=1$ SCFTs. These SCFTs are deformed by a VEV, flowing to a family a gapped QFT in $(2+1)$-dimensions. This is captured by a corresponding deformation in the gravity side of the duality, that ends smoothly.}
 \item{We calculated Wilson loops using the holographic description, by minimizing the action of a fundamental string.} 
 \item{Studying the energy of the quark-antiquark pair $E$ in terms of their separation $L$, we found a law that encompasses all of our numerical experiments. This law reads.
 \begin{equation}
 E(L)= -\frac{a}{L} +\frac{\gamma}{b}(1- e^{- b L} ).
 \end{equation}
 This shows a conformal behaviour for small separations $E\sim -\frac{a}{L}$, followed by a confining behaviour for intermediate separations $E\sim \gamma L$, that finally, for large separations, is dominated by a screened behaviour, namely $E\sim \frac{\gamma}{b}$.
 }
 \item{There are various things that distinguish the screening we observe from the one found in canonical theories (like QCD). Most notably, the `dynamical quarks' need not to be associated with the node for which we compute the Wilson loop. See the explanation around eq.(\ref{operatorred}) and Figure \ref{fig:screeningquiver}. }
\item{The procedure to find the minimal action for the F1-string, the string profile, the energy $E$ and the separation $L$ consist in minimizing the action using `splines'. A discretisation procedure described in Section \ref{algor} and Appendix \ref{sec:RobinHood} is also used. We make the code used publicly available.}
\end{itemize}
Let us list some topics for further research.
\begin{itemize}
 \item{It is clear that rich phenomenology is hidden in the curves $E(L)$, plotted across the paper. It would be interesting to understand some special cases of this, for example what occurs in very symmetric quiver diagrams. Similarly, the phenomenology of the coefficients $|a|,|b|,|\gamma|$ seems to be rich, presenting dependence on the quiver, the length of it, certain hierarchies among those coefficients, etc.}
\item{It would be nice to apply similar (variational) techniques used here to calculate other observables, like `t Hooft loops, Entanglement Entropy, complexity, etc. These observables respond to an `action' of the form in eq.(\ref{NGgeneric}). In some of these cases, one may need to deal with PDEs.}
\item{It is of interest to study the fluctuations of the string in the directions of the $S^2\times \Sigma_2$ part of the space. We believe these fluctuations will show stability of our probes, but it is worth checking that. We need to make sure that we are exploring the lowest possible energy configuration. The code can efficiently work in these situations.}
\item{It is also of interest to resolve directly the Euler-Lagrange equations (\ref{eqB3})-(\ref{bcs}) using the usual software, to cross-check our results.}
\item{It would be interesting to apply the techniques in this paper to construct flows between CFTs in dimension $D$ and gapped QFTs in dimension $(D-1)$. After this, calculating Wilson loops in these models following this work. This could be done for the SCFT-models in \cite{Macpherson:2024frt},\cite{Akhond:2022awd},\cite{Akhond:2022oaf},\cite{Legramandi:2021aqv},\cite{Akhond:2021ffz},\cite{Lozano:2020bxo},\cite{Lozano:2020txg}. Similarly, it would be interesting to see how our techniques apply to the case of a flow between CFTs in different dimension.}
\end{itemize}
Clearly, this paper opens new paths to explore. We believe it would be important to have various independent checks of our results.

\section*{Acknowledgments} For discussions, comments on the manuscript and for sharing their ideas with us, we wish to thank: Jerem{i}as Aguilera-Damia, Francesco Bigazzi, Nikolay Bobev, Aldo Cotrone, S. Prem Kumar, David Mateos, Simon Ross, Javier Subils. We are supported by the grants ST/Y509644-1, ST/X000648/1 and ST/T000813/1. The work of AF has been supported by the STFC Consolidated Grant ST/V507143/1 and by the EPSRC Standard Research Studentship (DTP) EP/T517987/1. The work of MG was funded by the European Union - Next Generation EU - National Recovery and Resilience Plan (NRRP) - M4C2 CN1 Spoke2 - Research Programme CN00000013 ``National Centre for HPC, Big Data and Quantum Computing'' - CUP B83C22002830001.


{\bf Open Access Statement}---For the purpose of open access, the authors have applied a Creative Commons Attribution (CC BY) licence to any Author Accepted Manuscript version arising. The \texttt{RobinHood} module is under a MIT licence.

\appendix

\section{Numerical Optimization with \texttt{RobinHood}}\label{sec:RobinHood}

To tackle the calculation of the Wilson loop of Section \ref{section-wilson}, in order to find the solutions presented in \ref{sec:numerical}, we use a numerical approach. We employed the Julia programming language \cite{Bezanson:2017julia}, creating a module called \texttt{RobinHood.jl}.\footnote{The module is publicly available on GitHub at \url{https://github.com/cu2mauro/RobinHood.jl}.} In this appendix, we illustrate how the module works, how it is constructed, and how to use it autonomously.

In the module, the problem is solved as such: the integration interval \texttt{I} of length \texttt{L} is created via the function
\begin{minted}{julia}
function interval(N::Int,L)
 I=Vector{Float64}(undef, 2N-1)
 I=range(0,(L/2)^2,length=N)
 I=sqrt.(I)
 I=[-I[end:-1:1];I]
 filter!(e->hash(e)!=hash(-0.0),I)
 return I
end
\end{minted}
The sub-intervals' length shrinks quadratically with \texttt{L} to better approximate the curved shape of the solution without needing \texttt{N} to be too large. The choice of it being quadratic is done after trying out the algorithm to numerically replicate the results of \cite{Maldacena:1997re} as a toy problem. 

The objective \texttt{action} function is a linear quadrature of the \texttt{lagrangian} function:
\begin{minted}{julia}
function lagrangian(x,r,rx,z,zx)
 L = @. NaNMath.sqrt(F2(r,z) + G2(r,z) * rx^2 + S2(r,z) * zx^2)
 return L
end
function action(c,I)
 hh=[I[2:1:end];0]-I
 pop!(hh)
 r=c[1:Int(length(c)/2)]
 z=c[Int(length(c)/2+1):end]
 bx=I[1:length(hh)]+hh./2
 rs=interpolate((I,), r, Gridded(Linear()))(bx)
 rx=diff(r)./hh
 zs=interpolate((I,), z, Gridded(Linear()))(bx)
 zx=diff(z)./hh
 Lag=lagrangian(bx,rs,rx,zs,zx)
 S=sum(hh .* Lag)
 return S
end
\end{minted}
To write them, we prioritize reusability, so that such an algorithm can work with more general Lagrangians in the future, and with higher-order splines if needed. 

Above, \texttt{c=[r;z]} is the array of optimization variables, and the \texttt{NaNMath.sqrt()} from the Julia package \texttt{NaNMath.jl} is employed because, during the optimization process, some guesses might go outside of the specified domain bounds. A DomainError in the optimization process would halt the execution, while a \texttt{NaN} value is simply discarded, forcing the optimizer to make a new guess.

The physical parameters $(P,N,z^*,q,\mu,l)$, together with a numerical cutoff $r_\infty$ and the $\alpha(z)$ of choice, are taken as inputs from a \texttt{config\_file.jl} or are read off a \texttt{background.jl} file. In particular, the values of $(P,z^*)$ and the choice of $\alpha(z)$ need to be fixed in the config file for the program to work.

The boundary conditions \eqref{bcs}, together with the domain bounds, are imposed via
\begin{minted}{julia}
cons(res, c, I) = (res .= [c[1], c[Int(end/2)], c[Int(end/2)+1], c[end]])
eqconst = [rinf, rinf, zstar, zstar]
lbounds = [fill(rstar,length(I));fill(0,length(I))]
ubounds = [fill(rinf,length(I));fill(P,length(I))]
\end{minted}
and used to define the OptimizationProblem to be solved with the \texttt{IPNewton} algorithm of \texttt{Optim.jl} \cite{optim}:
\begin{minted}{julia}
optprob = OptimizationFunction(action, Optimization.AutoReverseDiff(true), 
 cons = cons)
prob = OptimizationProblem(optprob, c0, I,; lcons = eqconst, ucons = eqconst, 
 lb = lbounds, ub = ubounds)
sol = solve(prob, IPNewton(),g_tol=1e-12,x_tol=1e-4)
\end{minted}
Here, \texttt{c0=[r0;z0]} is the initial guess of the optimization process. Our algorithm is robust enough to find the same solution for a wide range of initial guesses. However, probably because of the nonlinear nature of the problem, if the initial guess is very noisy, some of the optimizations do not converge. This is an issue that can be likely solved by a thoughtful choice for the optimization parameters, which we are planning to do in the future.

Multi-Threading parallelism is, for the present time, implemented via the \texttt{Threads.@threads} macro, yielding a simple but effective way to run the algorithm with different values of the parameters $(L,P,z^*)$ in a short time. We also plan to implement a more thoughtful parallelization, in case future analyses require more computing time.

Inside the module, we also include all the functions used for the plots of Section \ref{algor}. Their details can be found in the \texttt{plots.jl} file.

Practically, the only commands needed to run the code in a Julia window with \texttt{RobinHood.jl} already installed via the instruction in its README file, are\begin{minted}{julia}
julia> ] #enter package mode
pkg> add https://github.com/cu2mauro/RobinHood.jl/
julia> using RobinHood
julia> CreateConfigFile()
julia> include("config_file.jl")
\end{minted}
The only options to be modified in order to explore the various configurations presented in this paper are the parameters of the physical background and the choice of the $\alpha(z)$ function, inside the \texttt{config\_file.jl} text file.


\bibliographystyle{JHEP}
\bibliography{main.bib}

\end{document}